\documentclass[a4paper,fleqn,usenatbib]{mnras}

\usepackage{savesym}
\usepackage{amsmath}
\savesymbol{iint}
\usepackage{txfonts}
\restoresymbol{TXF}{iint}
\usepackage{longtable}
\usepackage{textcomp}
\usepackage{adjustbox}
 \usepackage{blindtext}

\usepackage{rotating,graphicx}

 \usepackage{epstopdf}
 \usepackage{epsfig}

\usepackage[T1]{fontenc}
\usepackage{ae,aecompl}

\usepackage{graphicx}	

\title[Ionized and molecular gas in nearby LIRGs]{Kinematics of the ionized and molecular gas in nearby luminous infrared interacting galaxies}

\author[Zaragoza-Cardiel et al.]{
Javier Zaragoza-Cardiel$^{1}$\thanks{E-mail: javier.zaragoza@astro.unam.mx},
John Beckman$^{2,3,4}$,
Joan Font$^{2,3}$,
Margarita Rosado$^{1}$,
\newauthor Artemi Camps-Fari\~na$^{2,3}$,
and Alejandro Borlaff$^{2,3}$
\\
$^{1}$Instituto de Astronom\'ia, Universidad Nacional Aut\'onoma de M\'exico, 04510, D. F., M\'exico\\
$^{2}$Instituto de Astrof\'isica de Canarias, C/ V\'ia L\'actea s/n, 38205 La Laguna, Tenerife, Spain\\
$^{3}$Department of Astrophysics, University of La Laguna, E-38200 La Laguna, Tenerife, Spain\\
$^{4}$CSIC, 28006 Madrid, Spain\\
}

\date{Accepted XXX. Received YYY; in original form ZZZ}

\pubyear{2016}

\begin{document}
\label{firstpage}
\pagerange{\pageref{firstpage}--\pageref{lastpage}}
\maketitle

\begin{abstract}
We have observed three luminous infrared galaxy systems (LIRGS) which are pairs of interacting galaxies,   with the Galaxy H$\alpha$ Fabry-Perot 
system (GH$\alpha$FaS) mounted on the 4.2m William Herschel Telescope at the Roque de los Muchachos Observatory, and combined 
the observations with the Atacama Large Millimeter Array (ALMA) observations of these systems in CO emission to compare the physical 
properties of 
the star formation regions and the molecular gas clouds, and specifically the internal kinematics of the star forming regions. 
We identified 88 star forming regions in the H$\alpha$ emission data-cubes, 
and 27 molecular cloud complexes in the CO emission data-cubes. The surface densities of the star formation rate 
and the molecular gas are significantly higher in these systems than in non-interacting galaxies and the Galaxy, 
 and are closer to the surface densities of the star formation rate 
and the molecular gas of extreme star forming galaxies at higher redshifts.
The large values of the velocity dispersion also show the enhanced gas surface density. The 
HII regions are situated on the ${\rm{SFR}}-\sigma_v$ envelope, and so are also in virial equilibrium.
Since the virial parameter decreases with the 
surface densities of both the star formation rate and the molecular gas, 
we claim that the clouds presented here are gravitationally dominated rather than being in equilibrium with the external pressure. 

\end{abstract}

\begin{keywords}
 galaxies: interactions -- galaxies: star formation -- galaxies: 
starburst -- galaxies: kinematics and dynamics -- 
galaxies: ISM 

\end{keywords}



\section{Introduction}

In the standard and simplified picture of galaxy evolution, the evolution can be driven by internal or external, and fast or slow 
processes. In early times, the evolution was dominated by fast and external processes such as mergers of galaxies \citep{2013seg..book....1K}.
Galaxy mergers are steps in the construction of more massive galaxies in the hierarchical model of galaxy formation which is commonly used within the 
standard cosmological model \citep{2005Natur.435..629S,2011EAS....51..107B}. 

In fact, the most intense star forming galaxies observed are the submillimiter galaxies observed at redshift 2-3 \citep{2005ApJ...622..772C} and 
$\rm{SFRs}\sim10^3\rm{M_{\odot}/yr}$ \citep{2010A&A...514A..67M}.
The high observed SFRs imply the important role that these galaxies play in the formation of massive galaxies. However, resolving the star formation on subkpc scales in these violent 
environments is not possible. Previous studies have used nearby analogs to connect their conclusions with high redshift galaxy evolution. For example 
a selected sample of local galaxies with high SFRs derived using H$\alpha$ \citep{2014MNRAS.437.1070G}, or selected samples of galaxy mergers since they were more common in the past.

One way to select galaxies to study the violent and fast evolution of galaxies at high redshifts is to use infrared luminosity. This selection criterion leads to the definition of 
luminous and ultraluminous infrared galaxies (LIRGs, $L_{\rm{IR}}>10^{11}\thinspace\rm{L_{\odot}}$; ULIRGs, $L_{\rm{IR}}>10^{12}\thinspace\rm{L_{\odot}}$) which 
are driven mostly by mergers in the nearby Universe, and are more common at higher redshifts \citep{1998ApJS..119...41K}.  The studies  
of the IR luminosity function in different redshifts made by \citet{2005ApJ...632..169L,2005ApJ...630...82P,magnelli11,magnelli13} find that 
these galaxies dominate the star formation activity for $z\gtrsim0.7$,  
when the star formation rate was the highest.   
The study of the frequency of mergers in 
these galaxies is difficult since the classification is morphological in observations of galaxies hardly resolved. Thus, depending on the method used different results arise, whether 
gas-rich mergers are a possible precursor of massive elliptical galaxies \citep{2010ApJ...721...98K} and then play a major role in the higher SFR at higher redshifts, or in situ 
star formation instead of mergers is the main driver of the higher SFR at higher redshifts \citep{2007A&A...468...33E}. Although the role of major mergers is not clear in the higher SFR densities, 
they are the main contributors of the morphological transformation and the quenching of the star formation from the main sequence of star forming galaxies to the quiescent 
population of galaxies \citep{2011ApJ...739L..40R,2011ApJ...742...96W}.

Nearby (U)LIRGs have received considerable
 study in detail. They have higher numbers of star forming clouds compare to normal star forming galaxies, 
 and host the highest star forming clouds in the local 
Universe, with values of the SFR surface density up to $\Sigma_{\rm{SFR}}=100\thinspace \rm{M_{\odot}\thinspace yr^{-1}\thinspace kpc^{-2}}$, 
\citep{2006ApJ...650..835A,2012A&A...541A..20A,2016A&A...590A..67P}, and although the star formation is more concentrated 
in the centre, they also present high extended star formation activity \citep{2011A&A...527A..60R} while 
the morphology of the gas and the stars differs, reaching the distribution peaks separations up to $1.4\rm{kpc}$ \citep{2006ApJ...650..850G}.

\begin{figure*}
\centering
\includegraphics[width=0.9\linewidth]{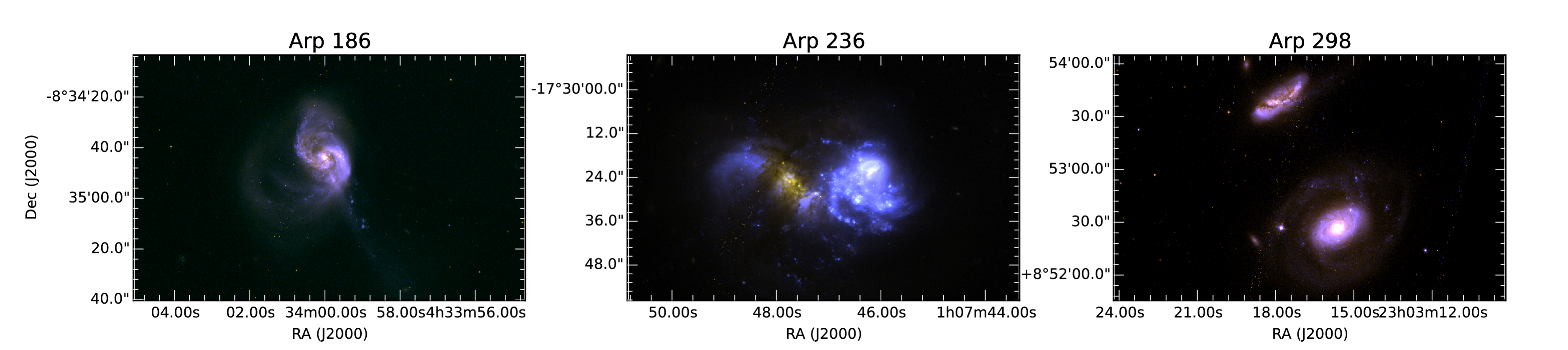}

\caption{Colour composite images of the three LIRGs using  F435W and F814W HST filters.}
\label{fig_color}
\end{figure*}

 The merger system the Antennae galaxies has an IR luminosity of $L_{\rm{IR}}=10^{10.83}\rm{L_{\odot}}$ \citep{2003AJ....126.1607S}, so it is not classified as a LIRG. However, given 
its distance of $D=22\rm{Mpc}$ \citep{2008AJ....136.1482S} and its high IR luminosity, 
this system has received a lot of attention in order to study a nearby example of an extreme environment and the behaviour of star formation within the system. In fact, \citet{2012ApJ...750..136W} found 
two populations of molecular clouds with a clear dichotomy $10^{6.5}\rm{M_{\odot}}$ in the molecular gas mass function. \cite{zaragoza14} identified 142 giant molecular clouds and 303 HII regions, confirming the 
two populations of molecular clouds and comparing them with two populations of HII regions which they observed. We will include the analysis of the Antennae galaxies as a local analog to the 
extreme environments at higher redshifts together with the analysis of three LIRGs. Since the Antennae galaxies are almost a LIRG we will include them as such in the discussion and conclusion.

Here we present the study of the ionized and molecular gas in a further three  nearby LIRGs observed with the Fabry-Perot GH$\alpha$FaS 
(Galaxy H$\alpha$ Fabry-Perot System) which have also been observed with the ALMA (Atacama Large Millimeter/submillimeter 
Array) radio interferometer, and we compare these new results with the observations of the Antennae galaxies presented in \cite{zaragoza14}.

The galaxies are Arp 186, a strongly interacting pair of galaxies in the late stage of the merger, which have an IR luminosity of $L_{\rm{IR}}=4\cdot10^{11}\rm{L_{\odot}}$ 
and are at a distance $D=62.6\rm{Mpc}$ \citep{2003AJ....126.1607S}.
Arp 236, a pair of interacting galaxies with nuclei which are still separated,and where one of them is strongly obscured by dust whereas the other is not. 
Arp 236 has an IR luminosity of $L_{\rm{IR}}=4.5\cdot10^{11}\rm{L_{\odot}}$ and is at a distance $D=78.6\rm{Mpc}$ \citep{2003AJ....126.1607S}. 
Arp 298, a pair of interacting galaxies in an early stage of a merger, have  an IR luminosity of $L_{\rm{IR}}=3.9\cdot10^{11}\rm{L_{\odot}}$ and are at a distance $D=65.2\rm{Mpc}$ \citep{2003AJ....126.1607S}. 
We show  in Fig. \ref{fig_color} the colour composite maps using images through the F435W and F814W HST filters.

Given the high spatial and velocity resolution of GH$\alpha$FaS and ALMA, we will explore the connection between the high SFR, and molecular gas surface densities in HII regions and 
molecular clouds in LIRGs, compared them with normal star forming regions, and extreme star forming regions at higher redshifts. Also we will include the analysis of their internal velocity dispersions, 
in order to understand the driver of the higher SFR and gas densities observed in nearby LIRGs, which could be the driver of the higher SFR in past epochs.

 In section \S2 we present the observations, and in section \S3 we present the catalog of the properties of the star forming clouds. 
In section \S4 we present the results and the analysis of the identified star forming regions. In 
section \S5 we discuss the results and in section \S6 we present our conclusions.

\section{Observations}

\subsection{GH$\alpha$FaS observations}
We have observed Arp 186, Arp 236, and Arp 298 with GH$\alpha$FaS \citep{2008PASP..120..665H}, a Fabry-Perot Interferometer at the Nasmyth focus of the 4.2m WHT (William Herschel Telescope) at the Roque 
de los Muchachos observatory, La Palma, Spain. We selected these LIRGs because they have been already observed with ALMA, from which the data are publicly available. Therefore, we are able 
to compare the recent star formation from the H$\alpha$ emission with the distribution of the properties of the molecular clouds. In table \ref{tab_obs} we give details about the observations.

\begin{table*}
	\centering
	\caption{Observations log.  Distances are taken from \citet{2003AJ....126.1607S}, assuming $\Lambda$CDM cosmological model with 
	$H_0=75\thinspace\rm{km\thinspace s^{-1}\thinspace Mpc^{-1}}$, and a flat cosmology with $\Omega_M=0.3$ and $\Omega_{\Lambda}=0.7$. }
	\label{tab_obs}
	\begin{tabular}{ccccccc} 
  \hline
 Name &RA (J2000)&Dec (J2000) & Date &Exp time & Seeing & Dist \\
   &     (hh:mm:ss)         &     ($^{\circ}$ $\mathrm{\prime}$ $\mathrm{\prime\prime}$) & dd/mm/yyyy& (min) &(arcsec) & (Mpc)\\
  \hline
Arp 186       &     04:33:59.8 & -08:34:44 & 30/10/2014 & 132  & 1.0    &  62.6    \\
Arp 236      &     01:07:47.2 &  -17:30:25 & 31/10/2014 & 152  & 1.0    &  78.6    \\
Arp 298      &     23:03:16.8 &  +08:53:01 & 30/10/2014 & 152  & 1.0    &  65.2    \\

\hline
\end{tabular}
\label{table_obs}
\end{table*}

Since GH$\alpha$FaS is at the Nasmyth focus, but without an optical derotator which would have reduced the field diameter and jeopardized the light throughput, we corrected for field rotation using 
the technique described in \citet{2010MNRAS.407.2519B}. GH$\alpha$FaS has a circular field of view of 3.4', free spectral range of $8\rm{\AA}$ which 
corresponds to 390km/s with a velocity resolution of 8km/s (although for Arp 186 we were constrained to 16km/s for technical reasons), with pixel size of 0.2'' and seeing limited angular resolution. 

\begin{figure*}

\centering
\includegraphics[width=0.95\linewidth]{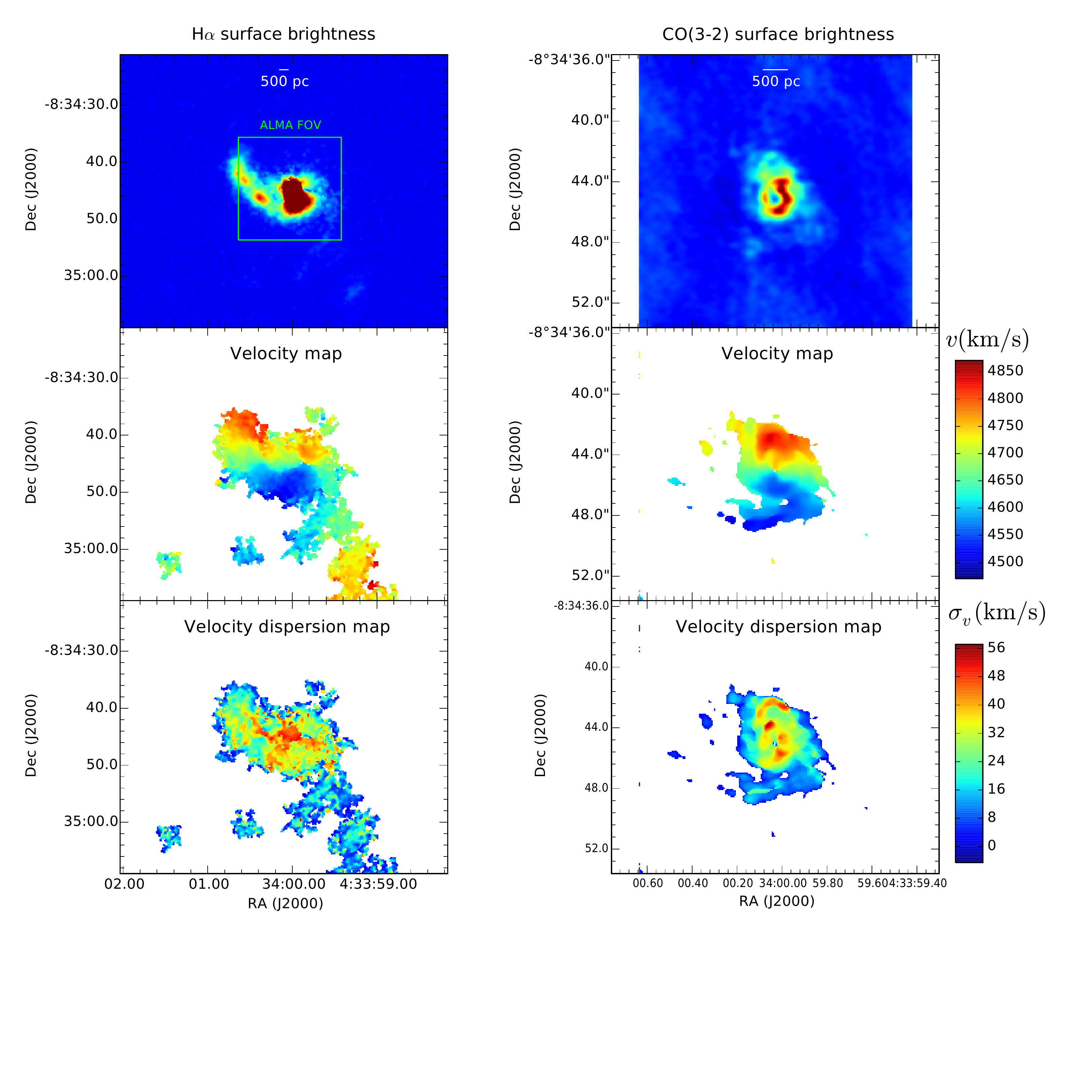}
%
%
%
%
\caption{Moment maps of Arp 186. (left) Derived from the GH$\alpha$FaS data cube. (right)
Derived from the ALMA observations \citep{2014ApJ...796L..15S}. 
Note that the effective field of view of ALMA is considerably smaller than that of GH$\alpha$FaS. 
The H$\alpha$ data cube as well as the product moment maps are available through CDS.}
\label{fig_maps1}
\end{figure*}

\begin{figure*}

\centering
\includegraphics[width=0.95\linewidth]{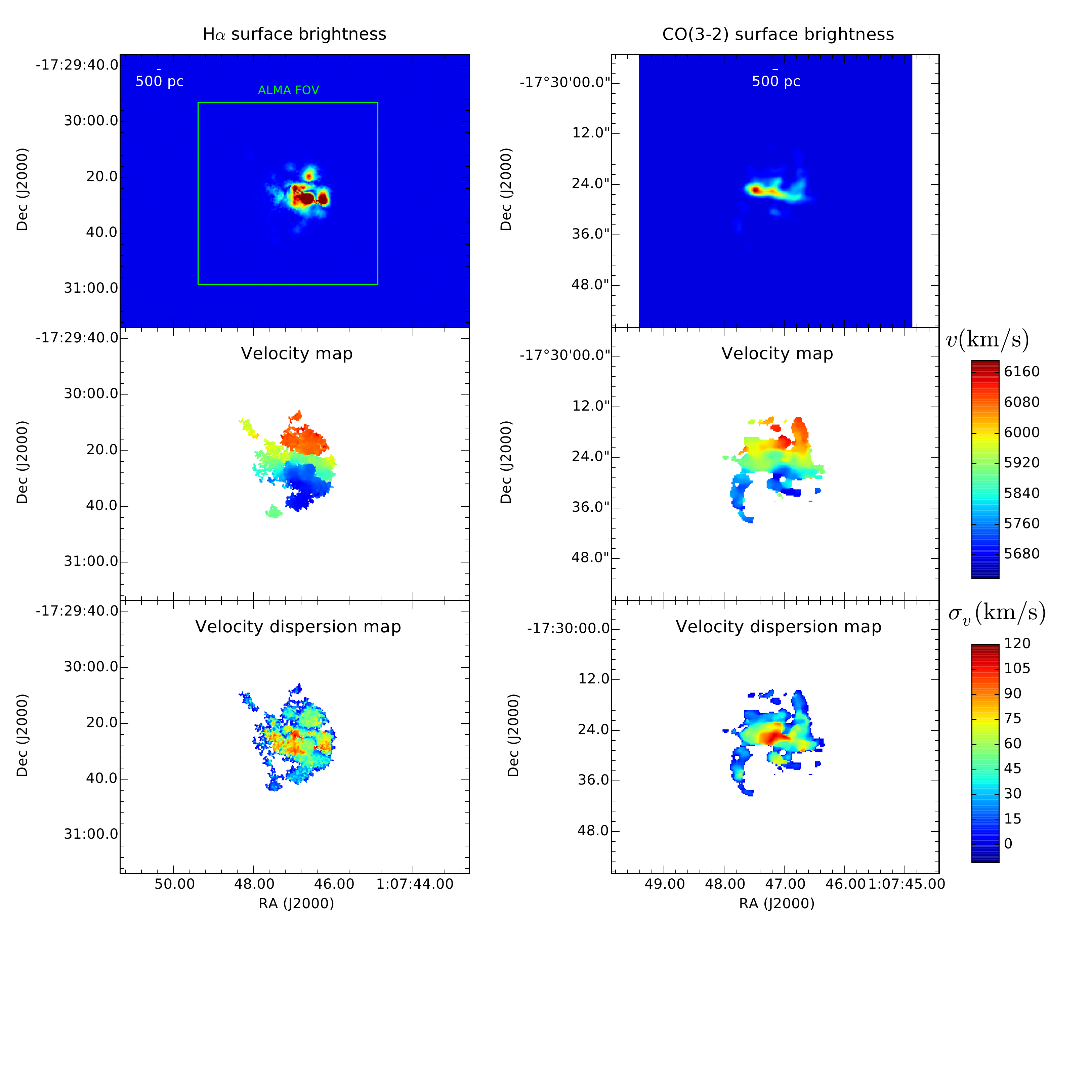}
%
%
%
\caption{Moment maps of Arp 236. (left) Derived from the GH$\alpha$FaS data cube. (right)
Derived from the ALMA observations \citep{2015ApJ...803...60S}. 
Note that the effective field of view of ALMA is considerably smaller than that of GH$\alpha$FaS. 
The H$\alpha$ data cube as well as the product moment maps are available through CDS.}

\label{fig_maps2}
\end{figure*}

\begin{figure*}

\centering
\epsfig{file=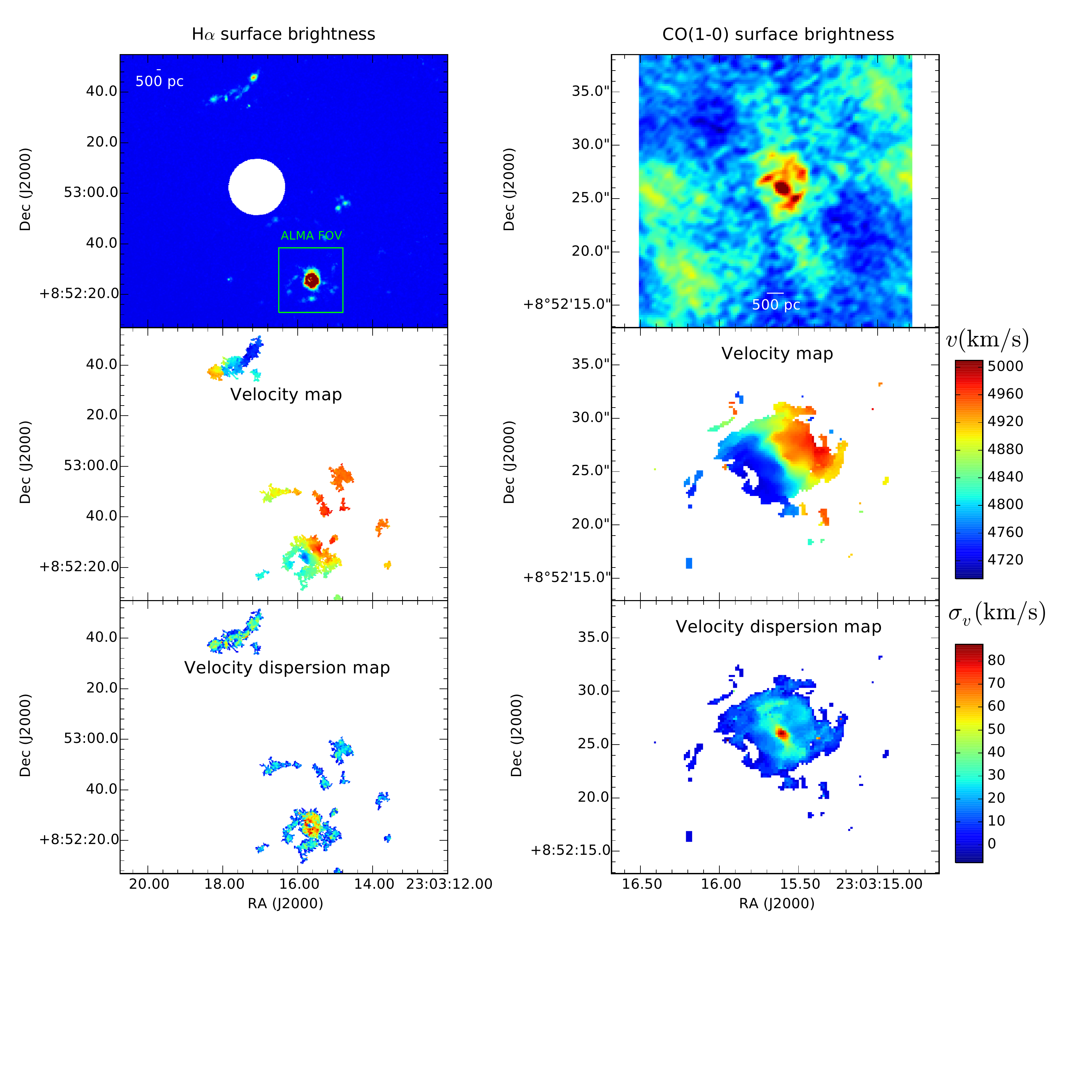,width=0.95\linewidth}
%
%
%
%
\caption{Moment maps of Arp 298. (left) Derived from the GH$\alpha$FaS data cube. (right)
Derived from the ALMA observations. 
Note that the effective field of view of ALMA is considerably smaller than that of GH$\alpha$FaS. 
The white circle in the H$\alpha$ surface brightness map masks a ghost (these artefacts may be present 
in Fabry-Perot observations, dependently on the source geometry).  
The H$\alpha$ data cube as well as the product moment maps are available through CDS.}
\label{fig_maps3}
\end{figure*}

After phase-correction, spectral smoothing, sky subtraction, and gaussian spatial smoothing, following 
the procedures detailed in \citet{2006MNRAS.368.1016D} we obtain the H$\alpha$ surface brightness, velocity, and velocity dispersion maps shown in 
Figs. \ref{fig_maps1}, \ref{fig_maps2}, \ref{fig_maps3} (left).

We use a continuum-subtracted and flux-calibrated ACAM (Auxiliary-port CAMera, \cite{2008SPIE.7014E..6XB}) 
H$\alpha$ image taken with the WHT to calibrate in flux the H$\alpha$ data-cubes following the procedure described in \citet{2012MNRAS.427.2938E}. 

The GH$\alpha$FaS data products: H$\alpha$ emission data cube, H$\alpha$ surface brightness, velocity dispersion, and velocity dispersion maps, are available through CDS.

\subsection{ALMA observations}

We used the ALMA observations of Arp 186 from \citet{2014ApJ...796L..15S},
the ones used here are the CO(3-2) emission, centred on the nuclear ring structure with the field of view seen in Fig. \ref{fig_maps1} (right), 
a spatial resolution of $0.45^{\rm{\prime\prime}}$, and 
a velocity resolution of $5\rm{km/s}$. 
In the case of Arp 236, we use the data from \citet{2015ApJ...803...60S} for the CO(3-2) emission, with the field of view seen in Fig. \ref{fig_maps2} (right), 
a spatial resolution of $0.45^{\rm{\prime\prime}}$, and a velocity resolution of $10\rm{km/s}$
Arp 298 ALMA observations are taken from the ALMA Early Science Cycle 2 project 2013.1.00218.S (Izumi et al., in preparation). We use in this paper the CO(1-0) emission which has the field of view centred on one galaxy 
of the merger, NGC 7469, seen in Fig.  
\ref{fig_maps3} (right). The spatial resolution is $0.4^{\rm{\prime\prime}}$. The reduced data in the archive has a resolution of $20\rm{km/s}$, although we repeated the reduction and 
were able to set the velocity resolution to $10\rm{km/s}$ 
in order to compare  better with the GH$\alpha$Fas observations. 

Due to the limitation imposed by the $u,v$ coverage, we miss the flux from scales larger than $53.75^{\rm{\prime\prime}}$ for Arp 186, $33.18^{\rm{\prime\prime}}$ for Arp 236, and $64.88^{\rm{\prime\prime}}$ for Arp 298. 
Considering the distances given in table \ref{table_obs}, we miss the flux from scales larger than $20\rm{kpc}$, $16\rm{kpc}$, and $13\rm{kpc}$, respectively.

The CO(3-2) (Arp 286 and Arp 236) and CO(1-0) (Arp298) surface brightness, velocity, and velocity dispersion maps for the three systems are shown in Figs. \ref{fig_maps1}, \ref{fig_maps2}, and \ref{fig_maps3} (right).  

\section{Properties of clouds}

We made use of \textsc{astrodendro}\footnote{\href{http://www.dendrograms.org}{http://www.dendrograms.org}}, 
a Python package to compute \textgravedbl dendrograms\textacutedbl of Astronomical data
\citep{2008ApJ...679.1338R} to extract the basic properties of the clouds (flux, radius, and velocity dispersion): the HII regions from the GH$\alpha$FaS data-cubes, and the 
molecular clouds from the ALMA data-cubes. 

The method is described in detail in \citet{2006PASP..118..590R,2008ApJ...679.1338R},  the application 
of the method to HII regions in interacting galaxies is shown in \citet{zaragoza15},  
while the comparison of the method between HII regions and molecular clouds is shown in \citet{zaragoza14}.

\subsection{Properties}

Once we have extracted the observable parameters, we can derive the properties of the clouds. 

\subsubsection{Properties of HII regions}

From the H$\alpha$ luminosity, $L_{\rm{H\alpha}}$, the size, $R$, and the velocity dispersion, $\sigma_v$, we can derive the 
star formation rate, SFR, using the calibration from \citet{1998ARA&A..36..189K}, $\rm{SFR}(\rm{M_{\odot}/yr})=7.9\cdot10^{-42}L_{\rm{H\alpha}}(\rm{erg/s})$. 
In order to compare with the properties of the HII regions found in the Antennae galaxies \citep{zaragoza14}, where the H$\alpha$ luminosities where corrected 
by dust attenuation, we correct using the average value found for $E(B-V)=0.4$ in the HII regions identified in the Antennae galaxies, and we applied the same correction 
to the HII regions from \citet{zaragoza15}. We show in 
\citet{zaragoza14} that the correction by dust attenuation does not affect the exponents of the scaling relations we will study here since the measured dust 
attenuation does not depend on the H$\alpha$ luminosity. Thus, a statistical study like this one is not affected by correcting all the regions by a constant 
value of dust attenuation. 
Then, we derive the SFR surface density, $\Sigma_{\rm{SFR}}=\frac{\rm{SFR}}{\pi R^2}$. 

Following \citet{relano2005} we also derive the electron density, $n_e$,  
assuming, as an approximation, spherical 
HII regions composed of hydrogen with uniform density \citep{1978spitzer}

\begin{equation}
\frac{L_{\rm H\alpha}}{\pi {R_{\rm{cm}}}^2}=h\nu_{\rm H\alpha}\alpha_{\rm H\alpha}^{\rm{eff}}(H_0,T)2.46\cdot 10^{17}\cdot n_e^2{R_{\rm{cm}}}
\label{eq:density}
\end{equation}

where $\alpha_{\mathrm H\alpha}^{\mathrm{eff}}(H_0,T)$ is the effective recombination coefficient of the H$\alpha$ emission
line, $h\nu_{\mathrm H\alpha}$ is the energy of an H$\alpha$ photon, and ${R_{\mathrm{HII}}^{\mathrm{cm}}}$ is the radius 
in cm of the HII region. The electron density is highly non homogeneous, so we note that the value estimated here is the square root of the mean squared 
electron density over the whole HII region. 

The ionized gas mass, $M_{\rm{HII}}$, is estimated using the $n_e$ estimated via Eq. \ref{eq:density}: 

\begin{equation}
M_{\mathrm{HII}}(\mathrm{M_{\odot}})  =  \frac{4}{3}\pi\thinspace R_{\mathrm{HII}}^3 \thinspace n_e \thinspace m_p 
= 1.57\times 10^{-17} \sqrt{L_{\mathrm H\alpha}\times R_{\mathrm{HII}}^3}
\label{eq2:mass}
\end{equation}

where $m_p=1.67\times 10^{-27}\mathrm{kg}$ is the proton mass, $L_{\mathrm H\alpha}$
is in erg/s, and $R_{\mathrm{HII}}$ is in pc. 

 We correct the measured velocity dispersion as explained in \citet{zaragoza14,zaragoza15}, assuming that the profiles are gaussians. 
Although \citet{2010MNRAS.407.2519B} showed that the H$\alpha$ profiles observed with Fabry-Perot are not well fitted by gaussians, their profiles 
where identified in the two dimensional H$\alpha$ images, and not in the datacubes. 
We checked carefully the degree of departure of our line profiles from gaussianity while preparing \citet{zaragoza15} 
and found that the effects were minimal. This is because astrodendro uses the line profiles to detect individual HII regions using the 
line profiles as well as the extent of a region, and the line wings (where departures from gaussianity are significant) 
are automatically rejected in the process.

We will use the ionized gas mass, and the corrected velocity dispersion to estimate the virial parameter, $\alpha_{\rm{vir}}= 5 \frac{\sigma_{v}^2 R_{\mathrm{HII}}}{GM_{\mathrm{HII}}}$, which 
is the ratio of the kinetic and the gravitational energy \citep{1992ApJ...395..140B}. This ratio is a measurement of what proportion of the velocity dispersion of the 
HII regions is due to their self gravity. We know 
that using the ionized gas mass instead of the total gas mass does not give us the actual value of the virial parameter but an approximately constant fraction of it,  even in the extreme case 
of the Antennae galaxies 
\citep{zaragoza14}. These roughly approximately constant factors are expected to vary from normal star forming regions to extreme high star forming regions. Therefore, the virial parameter for 
the HII regions presented here is a very approximated one, although the uncertainties in the virial parameter are more affected by deviation 
from sphericity and homogeneous density \citep{1992ApJ...395..140B}.

\subsubsection{Properties of molecular clouds}

In the case of molecular clouds we measure directly from the data-cubes the CO flux, $F_{\mathrm{CO}}$, the radius of the molecular cloud, 
$R_{\mathrm{CO}}$, and its velocity dispersion, $\sigma_v$. From the CO flux we can estimate the CO luminosity, $L_{\rm{CO}}$, following \citet{1992ApJ...398L..29S}:

\begin{equation}
\begin{array}{ll}
\frac{L_{\mathrm{CO}}}{\mathrm{K}\thinspace \mathrm{km}\thinspace \mathrm{s}^{-1}\thinspace \mathrm{pc}^2} & =
3.25\thinspace \times 10^7 \left(\frac{\nu_{\mathrm{rest}}}{\mathrm{GHz}}\right)^{-2}(1+z)^{-1}\\
& \times \left(\frac{D_L}{\mathrm{Mpc}}\right)^2\left(\frac{F_{\mathrm{CO}}}{\mathrm{Jy}\thinspace \mathrm{km}\thinspace \mathrm{s}^{-1}}\right)
\end{array}
\label{lumco}
\end{equation}

where $\nu_{\mathrm{rest}}$ is the rest frequency of the line ($345.796\thinspace \mathrm{GHz}$ in the case of CO(3-2), and $115.270\thinspace \mathrm{GHz}$
in the case of CO(1-0)), $D_L$ is the
luminosity distance, and $F_{\mathrm{CO}}$ is the velocity-integrated flux measured
in the data cube. 

Then, the molecular gas, $M_{\rm{mol}}$ mass is estimated from the CO luminosity:

\begin{equation}
 M_{\mathrm{H_2}}=\alpha_{\mathrm{CO}}\thinspace L_{\mathrm{CO}},
\label{eq:molmass}
\end{equation} 

where $\alpha_{\mathrm{CO}}=2m_{\mathrm{H}}X_{\mathrm{CO}}$, 
$m_{\mathrm{H}}$ is the mass of the Hydrogen atom, and 
$X_{\mathrm{CO}}$ is the empirical CO-H$_2$ conversion factor. We multiply by 1.8 the $\alpha_{\mathrm{CO}}$ factor for the use of CO(3-2) instead of 
CO(1-0) in Arp 186, and Arp 236, since taking the Antennae galaxies as a reference LIRG, the average ratio is $\frac{I_{CO}(1-0)}{I_{CO}(3-2)}\simeq1.8$ \citep{2012ApJ...745...65U}. 
 It is unclear what $\alpha_{CO}$ factor is convenient for LIRGs since this factor depends on the temperature, density, and metallicity. The range covers 
an interval between at least four times smaller and three times larger than the canonical Milky Way value, 
$\alpha_{CO}=4.8\rm{M_{\odot}(K\thinspace km\thinspace s^{-1})^{-1}}$, \citep{2012ApJ...751...10P,2013ARA&A..51..207B}.
We adopt the value of $\alpha_{CO}=4.8\rm{M_{\odot}(K\thinspace km\thinspace s^{-1})^{-1}}$, the same value as that adopted 
in similar studies of the Antennae galaxies \citep{2012ApJ...750..136W,2012ApJ...745...65U,zaragoza14}.
 Thus, our estimates of the molecular gas masses could be overestimated by a factor up to 4, or underestimated by a factor of 3, although the trends we find in this study would not change. 
We take into account the Helium and heavier elements in the molecular gas masses multiplying a factor 1.36, $M_{\rm{mol}}=1.36M_{\rm{H_2}}$ \citep{1973asqu.book.....A}.

Once we have the molecular gas mass, we derive the molecular gas mass surface density, $\Sigma_{\rm{mol}}=\frac{M_{\rm{mol}}}{\pi R^2}$.
We decided to study the surface density of the SFR and the molecular gas mass in order to remove as much as possible resolution effects since these LIRGs (except for the Antennae galaxies) are further away 
than the extragalactic sources we have used for comparison. 

We also estimate the virial parameter for the molecular clouds, $\alpha_{\rm{vir}}= 5 \frac{\sigma_{v}^2 R}{GM_{\rm{mol}}}$ to explore what fraction of the velocity dispersion is driven by the self gravity. 

\subsubsection{Uncertainties}

We followed the procedure (the bootstrapping method) explained in \citet{2006PASP..118..590R}
to derive the
errors in the measured parameters. This uncertainty does not include the intrinsic error 
of the flux in the data cubes. 

 The noise rms intensities in a single channel are  
$l_{\mathrm{rms}}=4\times10^{-18}\mathrm{erg\thinspace s^{-1}\thinspace cm^{-2}\thinspace \AA{}^{-1}}$ ($5.7\thinspace\mathrm{\mu Jy}$), 
$l_{\mathrm{rms}}=1\times10^{-18}\mathrm{erg\thinspace s^{-1}\thinspace cm^{-2}\thinspace \AA{}^{-1}}$ ($1.4\thinspace\mathrm{\mu Jy}$), and 
$l_{\mathrm{rms}}=3\times10^{-18}\mathrm{erg\thinspace s^{-1}\thinspace cm^{-2}\thinspace \AA{}^{-1}}$ ($4.3\thinspace\mathrm{\mu Jy}$), 
for the GH$\alpha$FaS observations of Arp 186 ($16\rm{km/s}$ channel separation), Arp 236 ($8\rm{km/s}$ channel separation), and Arp298 ($8\rm{km/s}$ channel separation) respectively. 
For the ALMA observations the noise rms intensities in a single channel are 
$l_{\mathrm{rms}}=1.7\thinspace\mathrm{mJy/beam}$ , 
$l_{\mathrm{rms}}=2.0\thinspace\mathrm{mJy/beam}$, and 
$l_{\mathrm{rms}}=1.1\thinspace\mathrm{mJy/beam}$, 
of Arp 186 ($5\rm{km/s}$ channel separation), Arp 236 ($10\rm{km/s}$ channel separation), and Arp298 ($10\rm{km/s}$ channel separation) respectively.

We identified and measured the properties of 23 HII regions in Arp 186, 46 in Arp 236, and 19 in Arp 298, and of 12 molecular clouds 
in Arp 186, 6 in Arp 236, and 9 in Arp 298.  The smaller number of molecular clouds compared to HII regions is because the field of view of the ALMA observations is smaller 
than the field of view of the GH$\alpha$FaS observations (Figs. \ref{fig_maps1}, \ref{fig_maps2}, \ref{fig_maps3}).

\begin{figure*}

   \centering
   \includegraphics[width=0.95\linewidth]{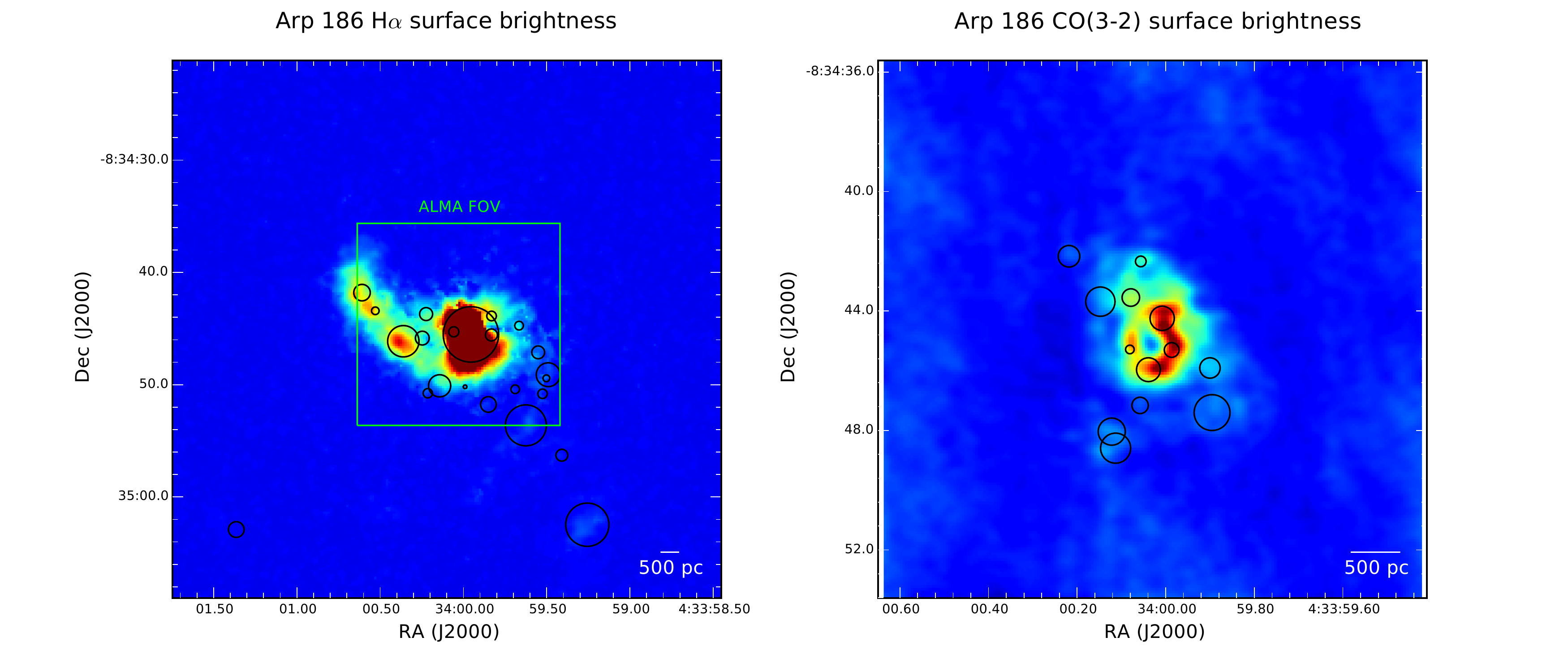}\\
   
   \includegraphics[width=0.95\linewidth]{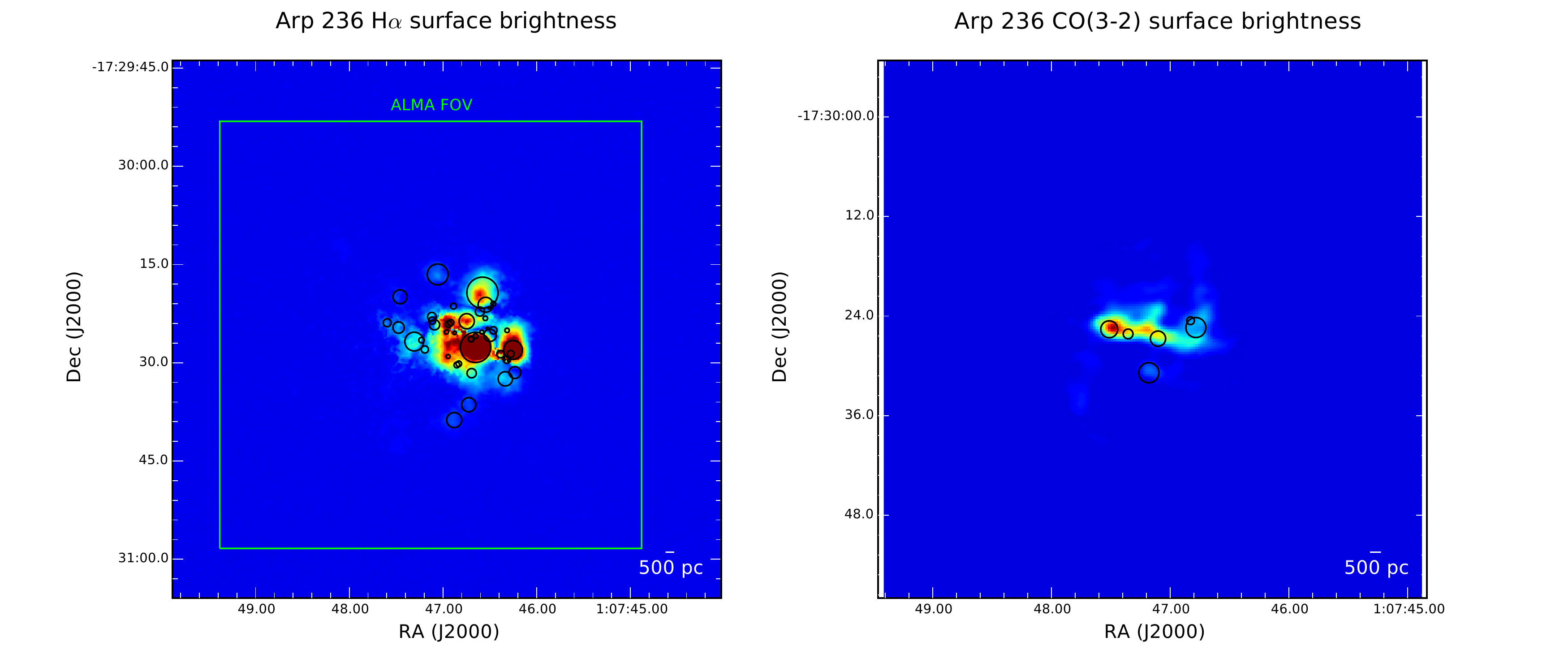}\\
   
   \includegraphics[width=0.95\linewidth]{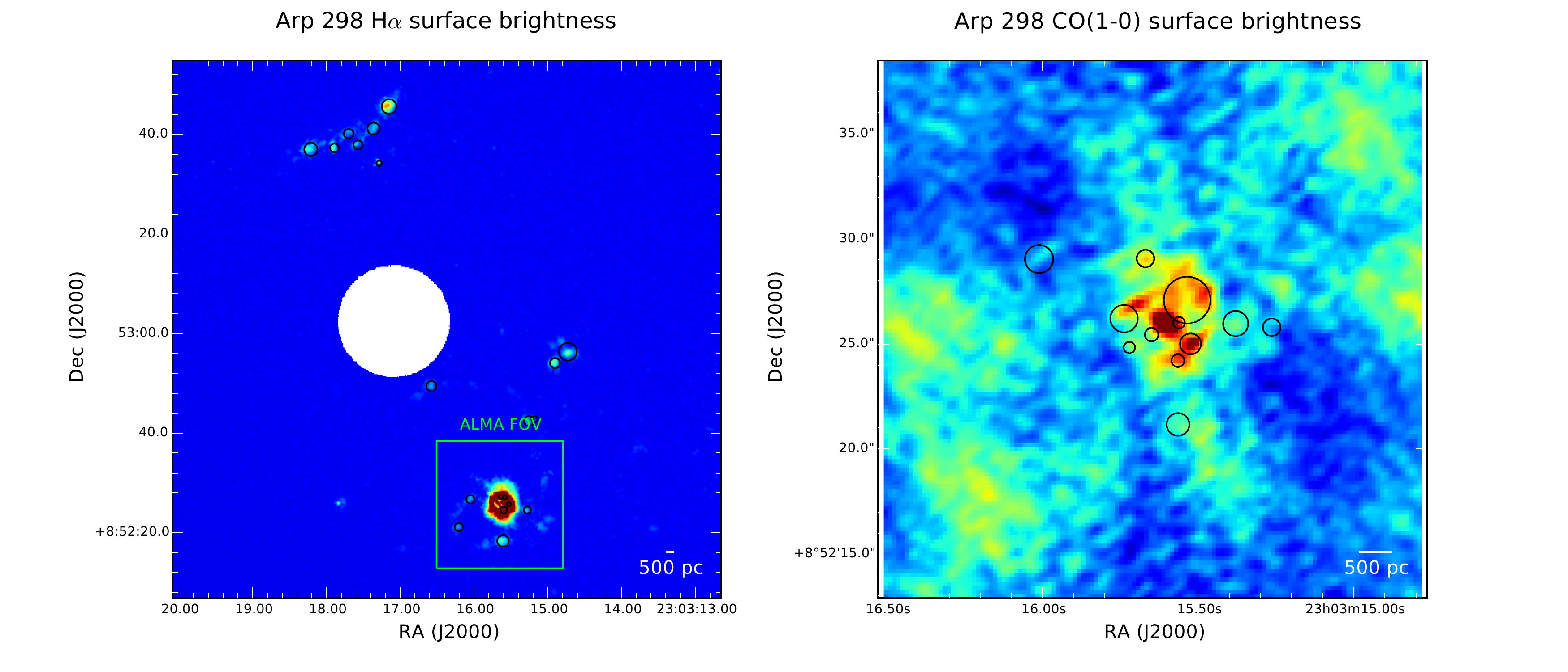}\\
   \caption{H$\alpha$ (left) and CO (right) surface brightness of Arp 186 (top), Arp 236 (middle), and Arp 298 (bottom). The circles 
   are the identified star forming regions.}
   \label{fig_loc}
      \end{figure*}

We show the properties measured and derived in Table \ref{table_hii}1 for the HII regions, in Table \ref{table_mol}2 for molecular clouds, 
 and the location of the regions in Fig. \ref{fig_loc}, where we can see that the centres of some of the regions do not coincide with 
the maximums of the surface brightness, due to the fact that the identification is made in the three dimensional datacubes, and these two centres may not coincide.

We have to clarify that when we talk about HII regions and molecular clouds we are talking about giant associations since at the distances of the galaxies under study these are regions and clouds which we can resolve; there are identified molecular clouds 
with radii up to 450 pc, and identified HII regions with radii up to 900 pc. We use the standard names for simplicity, and 
hereinafter we use the simplifying term clouds when we refer inclusively to molecular clouds and star forming regions in situations where we are incorporating results which are common to both species. 

%
%

\section{Results}

\begin{figure}

   \centering

   \includegraphics[width=0.9\linewidth]{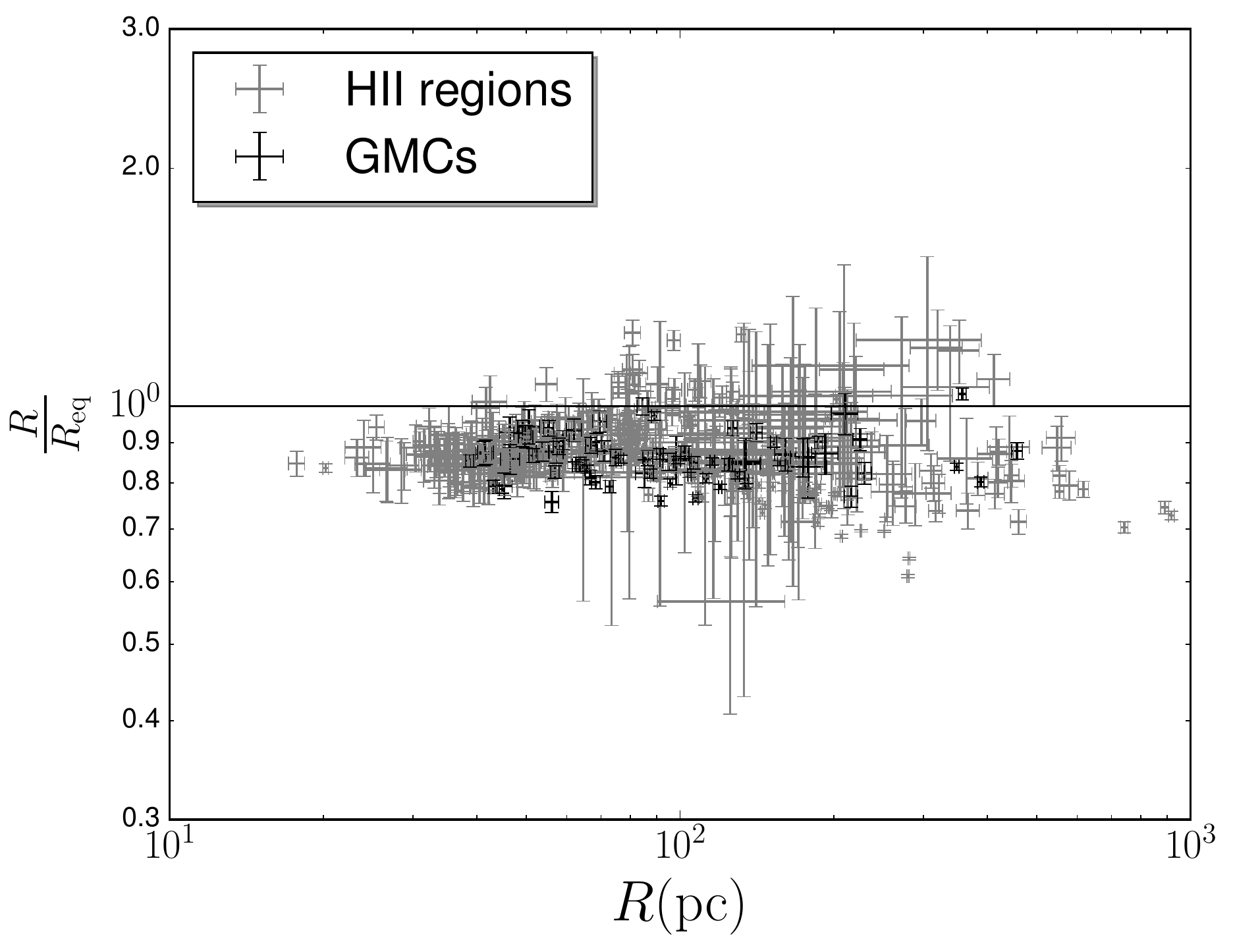}\\

   \caption{Ratio between the radius used in this work ($R$, described in \citet{2006PASP..118..590R}) and the equivalent radius ($R_{\rm{eq}}$, using the equivalent circular area), 
   $\frac{R}{R_{\rm{eq}}}$, versus the radius $R$ for the HII regions (in grey) presented here plus those of the Antennae galaxies from \citet{zaragoza14}, and for GMCs 
   (in black) presented here plus those of the Antennae galaxies from \citet{zaragoza14}.  The solid line represents the $\frac{R}{R_{\rm{eq}}}=1$ relation.}
   \label{rad_ratio}
      \end{figure}

Since we choose to use the method described in \citet{2006PASP..118..590R} to estimate the radius of the regions, in Fig. \ref{rad_ratio} we present the ratio 
of the radius estimated here over the radius estimated using the circular equivalent area ($\frac{R}{R_{\rm{eq}}}$). This area is defined by the circular area 
 equal to the exact area of the region, and then using this radius as the radius of the region, what we call $R_{\rm{eq}}$. This equivalent radius is also 
used in several catalogs of molecular clouds \citep{2009ApJ...699.1092H,2010ApJ...723..492R} and HII regions \citep{relano2005,gutierrez11,zaragoza13}. In Fig. \ref{rad_ratio} we show that the ratio between these two estimates does practically 
does not vary, with an average value of $R_{\rm{eq}}=0.86\pm0.14$. The use of different methods make difficult to compare between studies, however the data used here (except the high redshift one) are consistent in 
the estimation of the radius of the regions. For the HII regions we use the same estimation of the radius, while for the GMCs, we use for comparison the molecular clouds of the Galaxy 
from \citet{2009ApJ...699.1092H}, where they use the equivalent radius, although not much variation is expected from both estimations.

\subsection{$\Sigma_{\rm{SFR}}$ and $\Sigma_{\rm{mol}}$}

  \begin{figure*}

   \begin{tabular}{cc}
   \centering
   \includegraphics[width=0.45\linewidth]{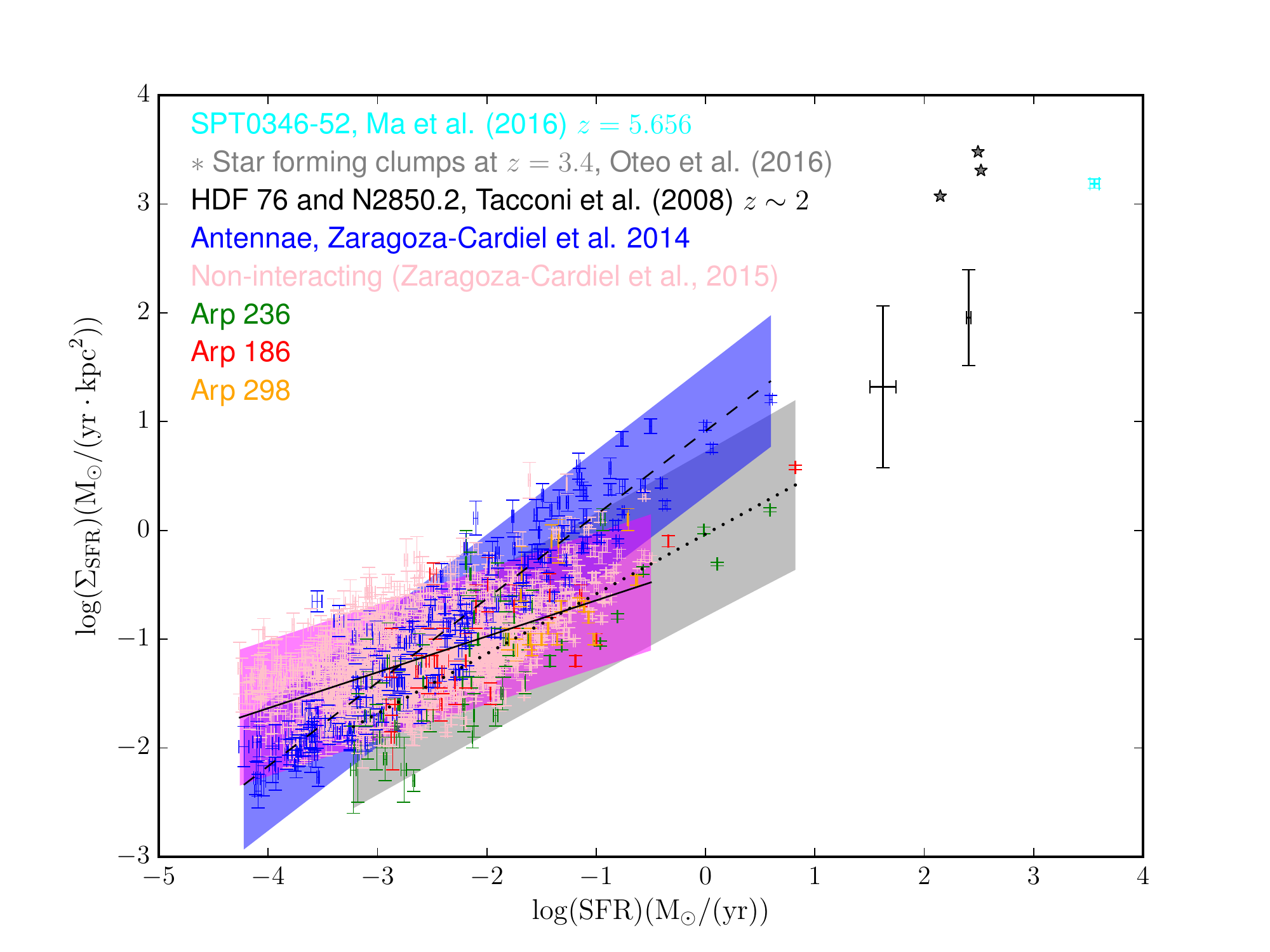}&
   \includegraphics[width=0.45\linewidth]{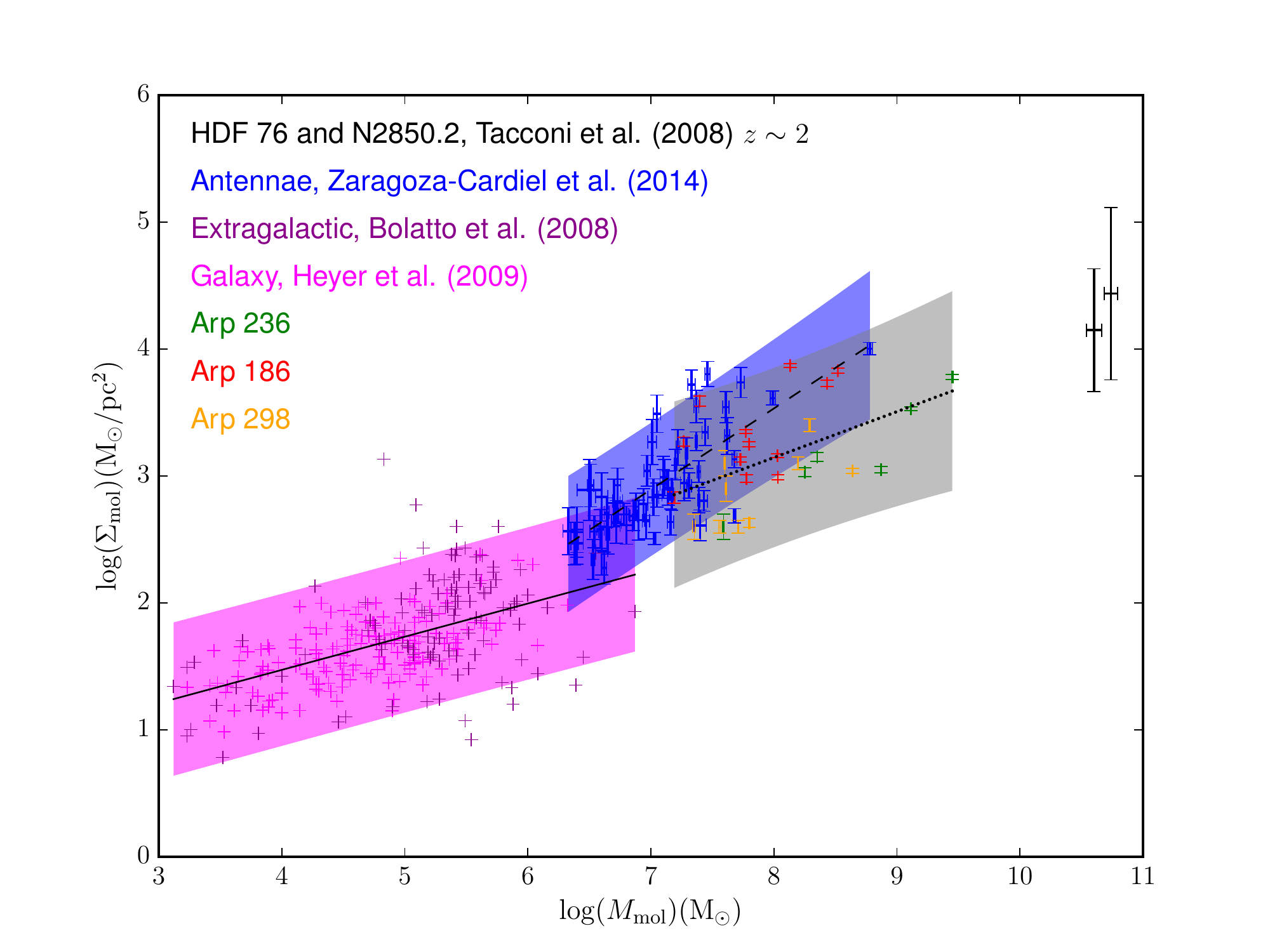}\\
   \end{tabular}
   \caption{Left: SFR surface density, $\Sigma_{\rm{SFR}}$, versus SFR, of the observed HII regions. 
   Right: molecular gas surface density, $\Sigma_{\rm{mol}}$, 
   versus molecular gas mass, $M_{\rm{mol}}$, of the observed molecular clouds. Solid lines are the fits for  regions
   in the non-interacting galaxy sample from \citet{zaragoza15}, and Galactic and extragalactic molecular clouds from 
   \citet{2008ApJ...686..948B,2009ApJ...699.1092H}, 
   while  dotted lines are the fits for the  regions in the LIRGS,  and dashed lines are for regions 
   in the Antennae. We show, as shaded areas (grey for LIRGs, blue for the Antennae, pink for the non-interacting galaxy and Galactic samples), 
   the 95\% confidence level that the true value of the ordinate is within the interval given by the value of the abscissa.}
   \label{fig_dens}
      \end{figure*}

We plot the SFR surface density, $\Sigma_{\rm{SFR}}$, versus the $\rm{SFR}$ in Fig. \ref{fig_dens} (left) for HII regions identified in the three systems presented here, 
plus the HII regions identified in \citet{zaragoza14} for the Antennae galaxies (blue). We also include, for comparison, the HII regions from \citet{zaragoza15} for 
a sample of non-interacting galaxies and two SMGs at redshift $z\sim 2$ from \citet{2008ApJ...680..246T}, three extreme star forming clumps at $z=3.4$ from \citet{2016arXiv160706464O}, 
and a strongly lensed dusty, star-forming galaxy at $z=5.656$ from \citet{2016arXiv160908553M}. 
 We fit separately radius error weighted linear fits to the $\log(\Sigma_{\rm{SFR}})=a\log\rm{SFR}+b$ relation for the HII regions in the LIRGS, the Antennae, and for those 
in the non-interacting galaxy sample. We weight 
by radius since it is the greatest source of uncertainty. 
 The results of the fits are in Table \ref{tab_fit_hii}. We estimate that the observed points are correlated using the Pearson Correlation Coefficient, and 
that the linear relation fits well enough through $\chi^2_{\rm{red}}$. 

\begin{table*}
	\centering
	\caption{Results of the linear fits to the $\log(\Sigma_{\rm{SFR}})=a\log\rm{SFR}+b$ relation for HII regions, and 
	$\log(\Sigma_{\rm{mol}})=a\log M_{\rm{mol}}+b$ relation for GMCs.}
	\label{tab_fit_hii}
	\begin{tabular}{ccccc} 
  \hline
  & a & b & \begin{tabular}{@{}c@{}}Pearson \\Correlation\\Coefficient \end{tabular}    & $\chi^2_{\rm{red}}$  \\

  \hline
LIRGs HII     &       $0.55\pm0.05$  & $-0.03\pm0.09$    & $0.78$ & $1.43$  \\
Antennae HII          &  $0.77\pm0.02$   & $0.91\pm0.06$  &$0.94$ &$1.45$   \\
Non-interacting HII    & $0.33\pm0.01$    & $-0.31\pm0.04$  &$0.58$ & $1.67$  \\

 \hline
LIRGs GMCs     &       $0.4\pm0.1$  & $0.2\pm1$    & $0.52$ & $4.65$  \\
Antennae GMCs           &  $0.64\pm0.07$   & $-1.6\pm0.5$  &$0.75$ &$0.86$   \\
Galaxy MCs \& normal extragalactic GMCs   & $0.26\pm0.03$    & $0.4\pm0.1$  &$0.52$ &  \\

\hline
\end{tabular}
\end{table*}

We show in Fig. \ref{fig_dens} (left) as a dotted line the fit to the LIRGs, as a dashed line the fit to the Antennae, and as a 
solid line the fit to the regions from non-interacting galaxies. We also show the 95\% 
confidence level that the true value of the ordinate is within the interval given by the value of the abscissa. We show 
these intervals as shaded areas, grey for LIRGs, blue for the Antennae, and pink for regions from non-interacting galaxies.



We observe a higher value for the  slope in the  $\log(\Sigma_{\rm{SFR}})=a\log\rm{SFR}+b$  
relation for the LIRGs than for the HII regions in the non-interacting galaxies. In fact, the values for $\Sigma_{\rm{SFR}}$ in the HII regions from the LIRGs 
are closer to the extreme $\Sigma_{\rm{SFR}}$ values found in SMGs (black, star-like,  and cyan points in Fig. \ref{fig_dens} left).
In the case of the Antennae galaxies the slope is even larger than for LIRGs, probably due to fact that is almost a LIRG and that it is at a 
closer distance. Therefore, the SFR surface densities in the Antennae are larger than the LIRGs because the resolution is finer \citep{2013ApJ...769L..12L,2016arXiv160607077L}.
The exponent of the HII regions in the 
non-interacting galaxies is in better agreement with the approximately constant H$\alpha$ surface brightness with radius found by 
\citet{1981MNRAS.195..839T,gutierrez11}, while the higher exponent of the HII regions in the LIRGs implies a higher SFR surface density 
in agreement with the triggered star formation regime found by \cite{zaragoza14,zaragoza15} in interacting galaxies. 

We show in Fig. \ref{fig_dens} (right) the molecular gas surface density, $\Sigma{\rm{mol}}$, versus the molecular gas mass, $M_{\rm{mol}}$, for 
molecular clouds identified in the CO emission of ALMA observations of the three systems presented here. Also we show the 
molecular clouds identified in the Antennae galaxies by \citet{zaragoza14} (blue) and molecular clouds from the Galaxy 
\citep{2009ApJ...699.1092H}, and extragalactic \citep{2008ApJ...686..948B}. We fit a radius error weighted linear fit to the $\Sigma{\rm{mol}}$-$M_{\rm{mol}}$ 
relation for the molecular clouds in the LIRGs,  in the Antennae, and those in the Galaxy-extragalactic separately.  The results of the fits are in Table \ref{tab_fit_hii}.
The $\chi^2_{\rm{red}}$ of the LIRGs in this case is higher than the rest because as we can see in Fig. \ref{fig_dens} (right) the errors are 
very low for the LIRGs, although still acceptable. We could not estimate the $\chi^2_{\rm{red}}$ for the clouds in the Galaxy and in external normal star forming galaxies 
since there are no uncertainties available. 

%
%


We show in Fig. \ref{fig_dens} (right) as a dotted line the fit to the LIRGs, as a dashed line the fit to the Antennae, and as a 
solid line the fit to the regions from non-interacting galaxies and the Galaxy. We also include the 95\% 
confidence level that the true value of the ordinate is within the interval given by the value of the abscissa. We show 
these intervals as shaded areas, grey for LIRGs, blue for the Antennae, and pink for regions from non-interacting galaxies and the Galaxy.
%

In the case of molecular clouds, we observe also that the slope in the 
$\log(\Sigma{\rm{mol}})=a\log M_{\rm{mol}}+b$ relation is greater for the molecular clouds in the LIRGs compare to the extragalactic and Galactic molecular clouds. 
As well as for the HII regions, the slope is even larger for the GMCs of the Antennae galaxies probably due to resolution effects in 
the GMCs of the LIRGs. However, the correlation coefficients for GMCs in LIRGs and in MCs in normal star forming galaxies show an increasing trend instead of 
a tight correlation. 
This indicates that 
there is a higher molecular gas density regime in the LIRGs with the highest values closer to the values of the two extreme star 
forming SMGs at redshift $z\sim2$, although the difference in the exponents is smaller, and the correlation more disperse, for the molecular clouds than in the HII regions.

\subsection{Velocity dispersion}

\subsubsection{$L_{\rm{H\alpha}}$-$\sigma_v$ envelope for HII regions}

  \begin{figure}
       \centering
   \includegraphics[width=0.9\linewidth]{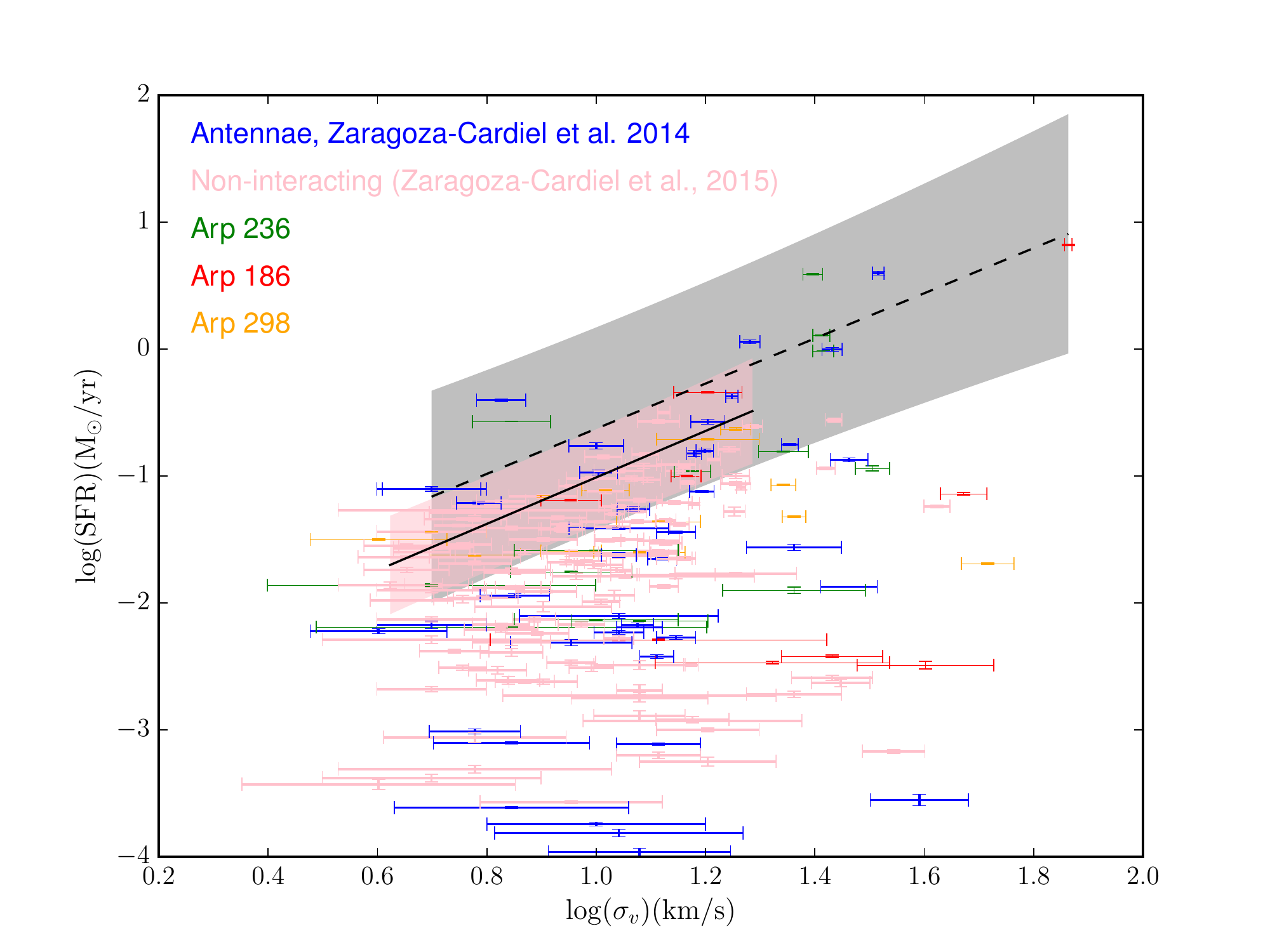}

   \caption{$\rm{SFR}$ versus the velocity dispersion, 
$\sigma_v$, of the observed HII regions. The solid line is the fit to the envelope of HII regions in the non-interacting galaxies sample, while 
the dashed line is the fit to the envelope of HII regions in LIRGs and the Antennae galaxies.
 We show, as shaded areas (grey for LIRGs+Antennae, pink for the non-interacting galaxy sample) ,
   the 95\% confidence level that the true value of the ordinate is within the interval given by the value of the abscissa, for the regions on the envelope.
} 
  
   \label{fig_sv_sfr}
      \end{figure}

Since we have the kinematical dimension from the data-cubes of GH$\alpha$FaS and ALMA, we can analyse the internal kinematics of the clouds by 
means of the velocity dispersion to see if the enhancement in the star formation rate is related to the velocity dispersion.

An envelope in the $L_{\rm{H\alpha}}-\sigma_v$ (equivalent to the one presented in Fig. \ref{fig_sv_sfr})  
has been observed by \citet{relano2005,2010MNRAS.407.2519B,zaragoza13}, and even used for 
distance estimates \citep{chavez,2015MNRAS.451.3001T}. The physical explanation for the cause of the envelope is 
that the regions on the envelope are 
virialized because they are gravitationally bound while the velocity dispersion of the regions far from the envelope is driven 
by stellar feedback as shown in \citet{relano2005,zaragoza13,zaragoza14,zaragoza15}. 
We show in Fig. \ref{fig_sv_sfr} the $\rm{SFR}$ versus the velocity dispersion, 
$\sigma_v$, of the observed HII regions. In the case of HII regions, the velocity dispersion is affected by stellar winds.  
 We use the method described in \citet{relano2005} to find the regions on the envelope, which consists in starting at a certain luminosity, choosing 
the region with the smallest velocity dispersion per logarithmic luminosity bin.
The results of the fit to the regions on the envelope  are in Table \ref{tab_fit_env}. In this case we fit together the Antennae galaxies with 
the LIRGs. We even could include the non-interacting galaxies sample since the envelope is the same.  

\begin{table*}
	\centering
	\caption{Results of the linear fits to the $\log(\rm{SFR})=a\log\sigma_v+b$ envelope for HII regions, and 
	the $\log(\Sigma_{\rm{mol}})$=$a\log\sigma_v+b$ relation for GMCs. }
	\label{tab_fit_env}
	\begin{tabular}{ccccc} 
  \hline
  & a & b & \begin{tabular}{@{}c@{}}Pearson \\Correlation\\Coefficient \end{tabular}    & $\chi^2_{\rm{red}}$  \\

  \hline
LIRGs+Antennae HII     &       $1.8\pm0.3$  & $-2.4\pm0.3$    & $0.89$ & $3.95$  \\
Non-interacting HII    & $1.8\pm0.2$    & $-2.8\pm0.2$  &$0.94$ & 1.40 \\

  \hline
LIRGs +Antennae GMCs     &       $1.2\pm0.2$  & $1.7\pm0.2$    & $0.57$ & $2.00$  \\
Galaxy MCs \& normal extragalactic GMCs   & $0.81\pm0.08$    & $1.38\pm0.04$  &$0.53$ &  \\

\hline
\end{tabular}
\end{table*}

%
%

 We show  
 the fit of the HII regions on the envelope of the LIRGs (including the Antennae) plotted as a dashed line in Fig. \ref{fig_sv_sfr}, while the fit of the regions on the envelope of 
 the non-interacting galaxy sample 
 is plotted as a solid line in Fig. \ref{fig_sv_sfr}. We include the 95\% confidence level , as explained in section \S4.1, for the regions on the envelope. We show 
these intervals as shaded areas, grey for LIRGs+Antennae, and pink for regions from non-interacting galaxies.


The  slopes in Table \ref{tab_fit_env} are consistent with those found previously for the $L_{\rm{H\alpha}}-\sigma_v$ 
relation by \citet{relano2005,2010MNRAS.407.2519B,zaragoza13,zaragoza14,zaragoza15}. 
\citet{zaragoza15} studied the properties of 1259, and 1054 HII regions in interacting and non-interacting galaxies, respectively. They found 
that the regions in more extreme environments such as merging galaxies are more likely to be gravitationally 
bound than the regions in non-interacting galaxies. Here we cannot perform such a statistical comparison because of the low number of HII regions in 
our sample of LIRGs. However, since the LIRGs here are merging galaxies and they are forming stars at a higher rate, we should expect that 
the regions here are also more likely to be gravitationally bound. The more dominated gravitationally a region is, the more the star formation depends on the 
internal velocity dispersion, as is the case for the regions on the envelope. We will discuss this in more detail using the virial parameter in the next section. 

  \begin{figure}
   \centering
   \includegraphics[width=0.9\linewidth]{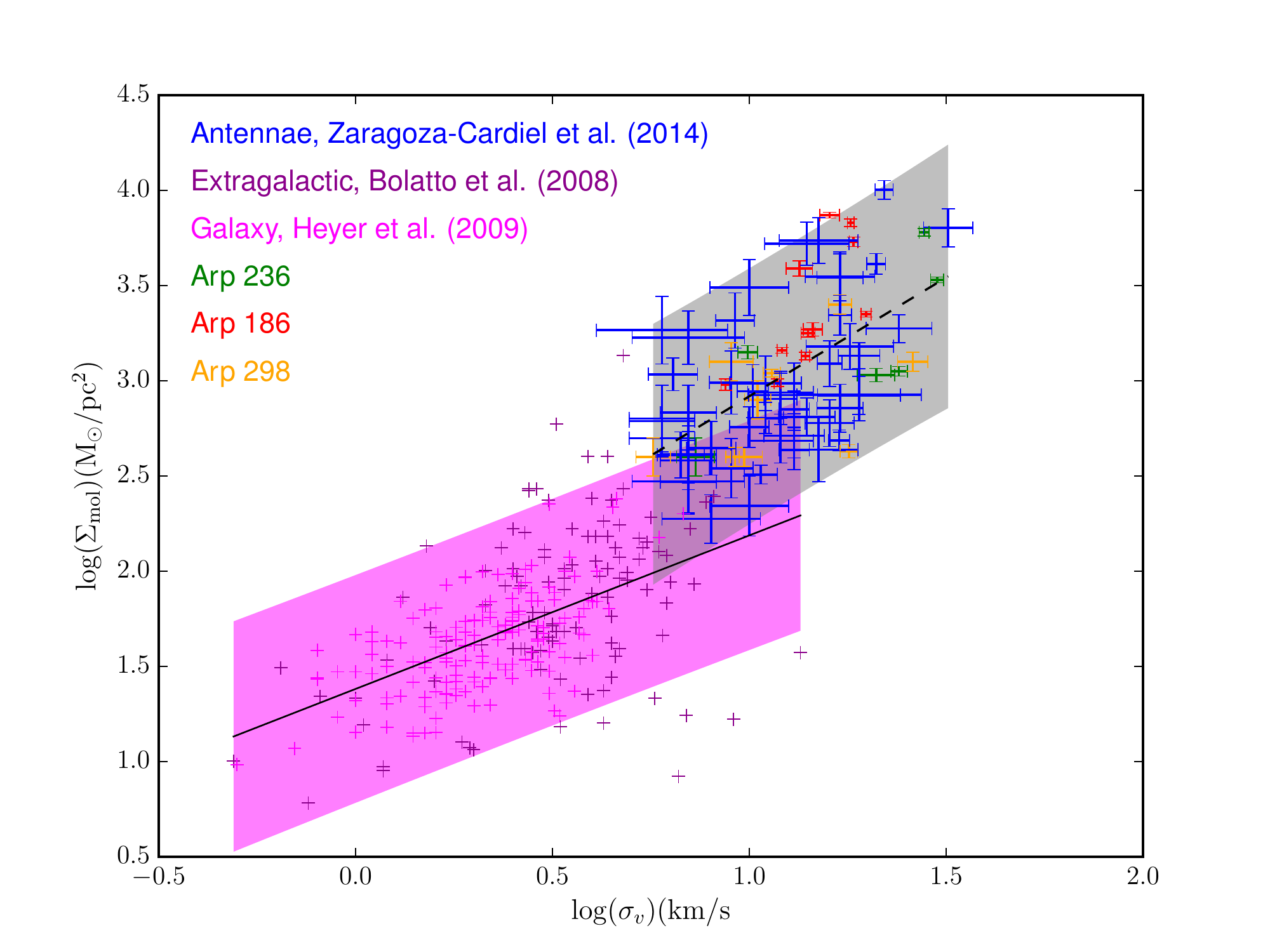}\\
   \caption{Molecular gas surface density, $\Sigma_{\rm{mol}}$, 
   versus the velocity dispersion, 
$\sigma_v$, of the observed molecular clouds.  The solid line is the fit to the $\Sigma_{\rm{mol}}$-$\sigma_v$ relation for 
molecular clouds in external normal star forming galaxies and in the Galaxy from \citet{2008ApJ...686..948B,2009ApJ...699.1092H}, while the dashed line is the fit to the 
molecular clouds in the LIRGs and the Antennae.
 We show, as shaded areas (grey for LIRGs+Antennae, pink for the normal star forming galaxy and Galactic samples),
   the 95\% confidence level that the true value of the ordinate is within the interval given by the value of the abscissa.
}
   \label{fig_sv_den}
\end{figure}

\subsubsection{$\Sigma_{\rm{mol}}$-$\sigma_v$ relation for molecular clouds}
      
We plot in Fig. \ref{fig_sv_den} the molecular gas surface density, $\Sigma_{\rm{mol}}$, 
   versus the velocity dispersion, 
$\sigma_v$, of the observed molecular clouds for the systems presented here and the Antennae galaxies (blue). 
Also we include for comparison Galactic molecular clouds (pink) from 
\citet{2009ApJ...699.1092H}, and extragalactic molecular clouds (grey) from non-interacting galaxies 
\citep{2008ApJ...686..948B}. Since the stellar feedback cannot act for the molecular clouds, the molecular gas surface density 
correlates with the velocity dispersion. 
We fit the $\Sigma_{\rm{mol}}$-$\sigma_v$ relation for the 
molecular clouds in the LIRGs (including the Antennae), and in the Galaxy and in normal star forming galaxies separately. 
  We also show in Fig. \ref{fig_sv_den} the 95\% confidence level, as explained in section \S4.1. We show 
these intervals as shaded areas, grey for LIRGs+Antennae, and pink for regions from normal star forming galaxies and the Galaxy.


 We show the results of the fits in Table \ref{tab_fit_env}. In the case of the clouds in the Galaxy and in external normal star forming galaxies  
we could not estimate the $\chi^2_{\rm{red}}$ 
since there are no uncertainties available. In Fig. \ref{fig_sv_den} we show the fit for the molecular clouds in the LIRGs and the Antennae plotted as a dashed line, and 
for molecular clouds in the Galaxy and in external normal star forming galaxies plotted as a solid line. 
We see that the slope in the $\log(\Sigma_{\rm{mol}})$-$\log\sigma_v$ is slightly higher in the case of molecular clouds 
in LIRGs.  However, the correlation coefficients in Table \ref{tab_fit_env} show that the increasing trend is disperse.  
For the highly turbulent molecular clouds in LIRGs, the molecular gas surface density seems higher per unit velocity dispersion, 
from which we infer that the turbulence permits the enhancement of the molecular gas density. 
Since the star formation depends on the molecular gas density, the star formation is also more efficient with increased turbulence.

  \begin{figure}

   \centering

   \includegraphics[width=1.0\linewidth]{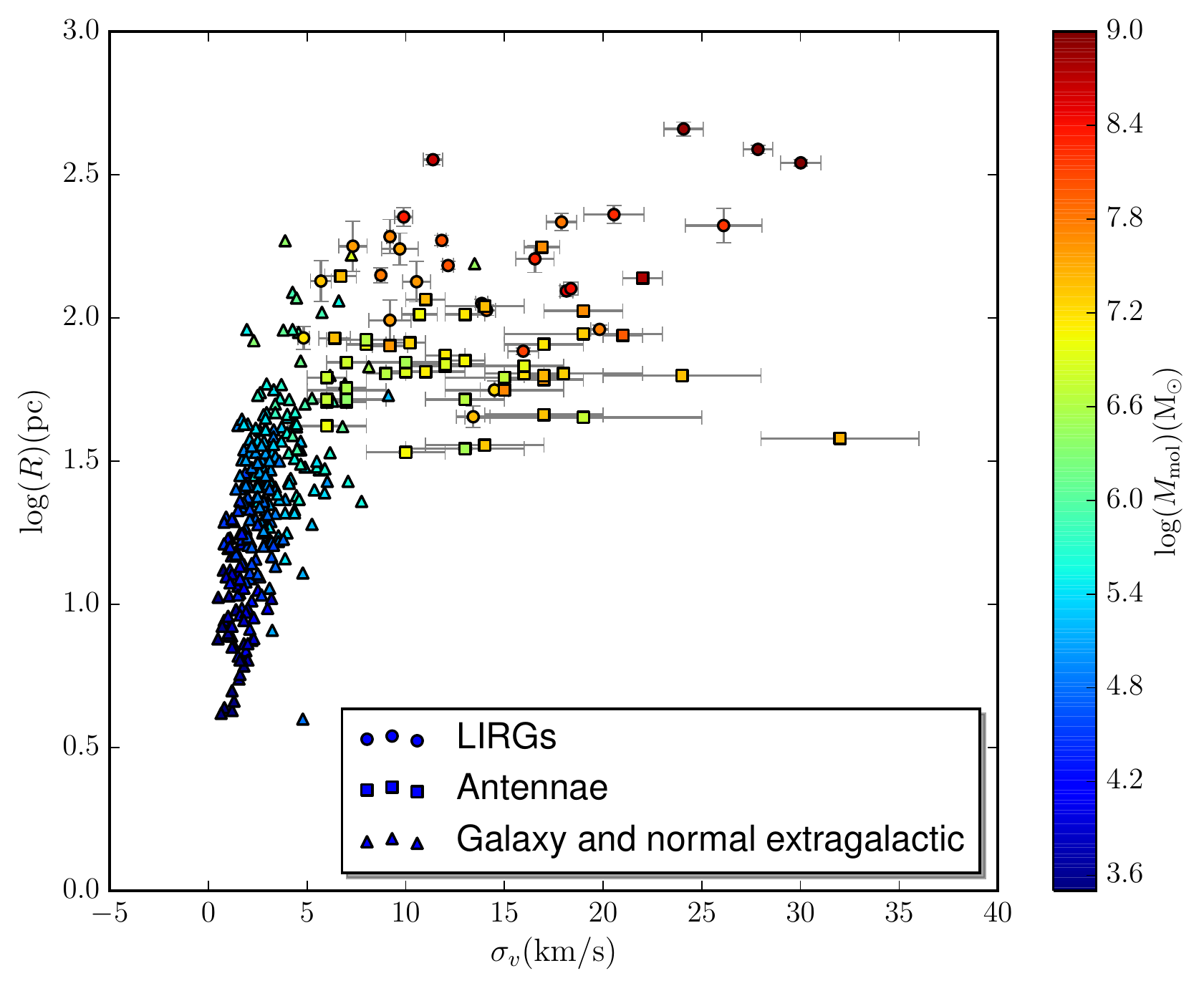}\\

   \caption{Radius, $R$, 
   versus the velocity dispersion, 
$\sigma_v$, of the observed molecular clouds, colour-coded with the molecular gas mass, $M_{\rm{mol}}$. 
Circles represent GMCs from LIRGs,  squares represent GMCs from the Antennae galaxies, while triangles represent MCs from the Galaxy and GMCs in 
external galaxies from \citet{2008ApJ...686..948B,2009ApJ...699.1092H}.}
   \label{fig_sv_rad}
      \end{figure}

In order to see if there is a typical scale where the turbulence enhances the molecular gas density, we plot in Fig. \ref{fig_sv_rad} 
the radius, $R$, versus the velocity dispersion, $\sigma_v$, of the molecular clouds in LIRGs (circles), the Antennae (squares),  and the 
molecular clouds in the Galaxy and in external galaxies (triangles) from \citet{2008ApJ...686..948B,2009ApJ...699.1092H}. The molecular clouds become 
more compact at radius $R\sim100 \rm{pc}$ since the radius stops to grow with the velocity dispersion, producing more massive molecular clouds. 
Notice that to see this density enhancement we have plotted the velocity dispersion, $\sigma_v$, and not the logarithm of it.

 \subsection{Virial parameter}
 
  \begin{figure*}

   \begin{tabular}{cc}
   \centering
   \includegraphics[width=0.45\linewidth]{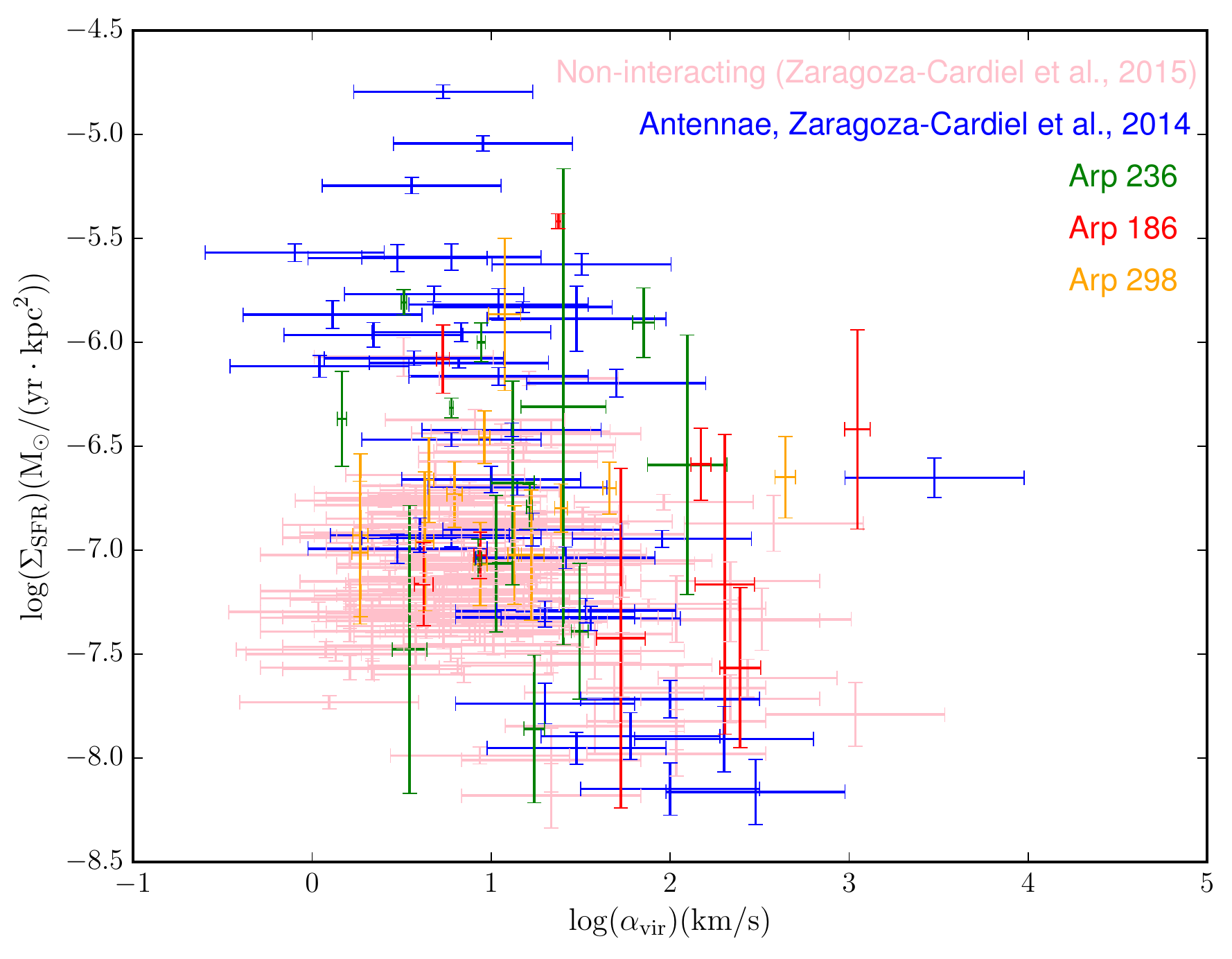}&
   \includegraphics[width=0.45\linewidth]{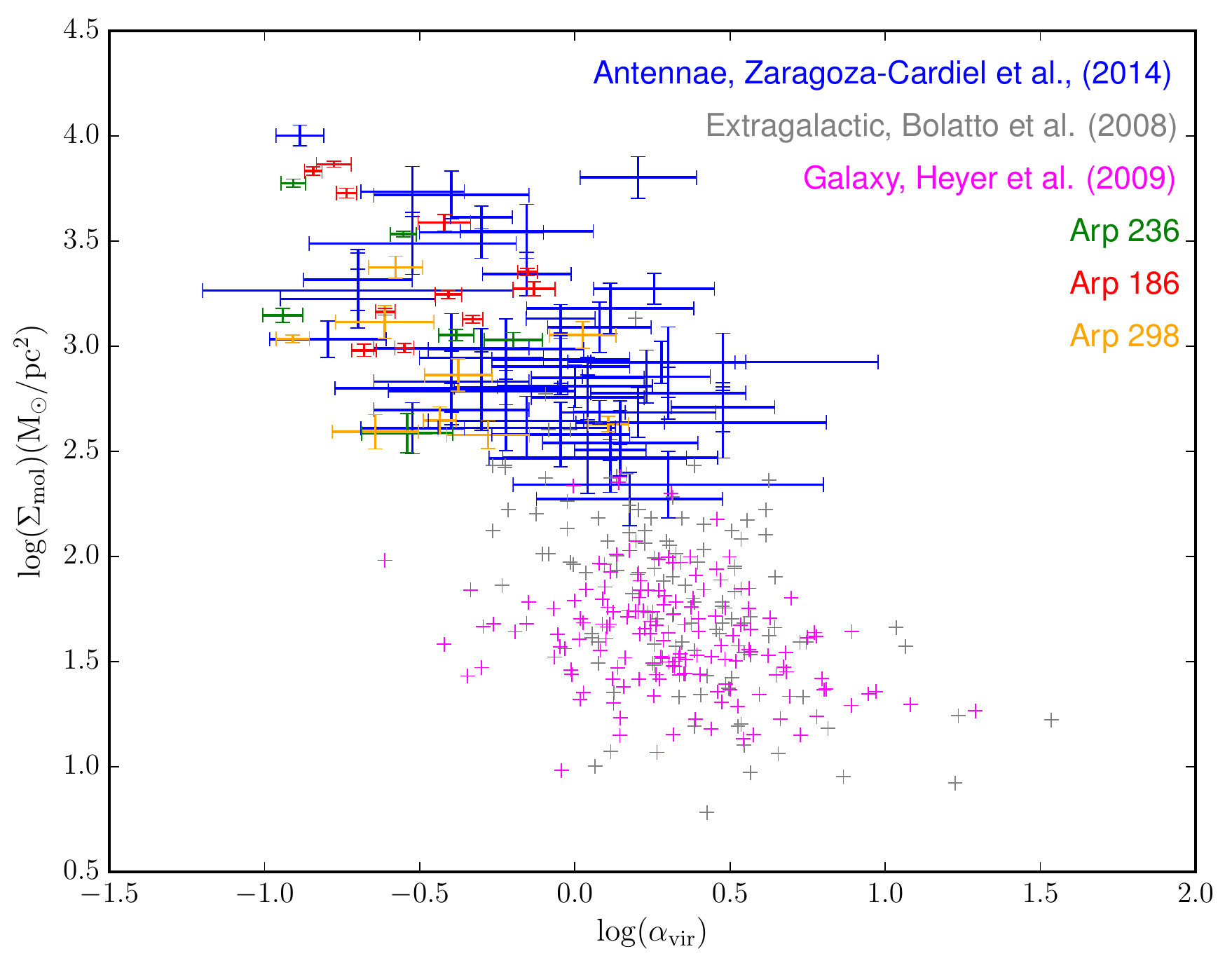}\\
   \end{tabular}
   \caption{Left: SFR surface density, $\Sigma_{\rm{SFR}}$, versus the virial parameter, $\alpha_{\rm{vir}}$, 
 of the observed HII regions. 
   Right: molecular gas surface density, $\Sigma_{\rm{mol}}$, versus 
  the virial parameter, $\alpha_{\rm{vir}}$, of the observed molecular clouds. }
   \label{fig_vir}
      \end{figure*}
 
 The force that makes the clouds denser is clearly gravity. We will show the importance of gravity in the massive clouds we have found in this work, as we did in 
 \citet{zaragoza14} for the Antennae galaxies. We use the virial parameter, $\alpha_{\rm{vir}}$, which is the ratio of the kinetic and gravitational energies, and gives
 us an estimate of to what extent the gravity is the driver of the velocity dispersion of the clouds. We estimate $\alpha_{\rm{vir}}$ as described in \S3. 
 We plot $\Sigma_{\rm{SFR}}$ versus $\alpha_{\rm{vir}}$ for the observed HII regions in Fig. \ref{fig_vir} (left) while we plot 
  $\Sigma_{\rm{mol}}$ versus $\alpha_{\rm{vir}}$ for the observed molecular clouds in Fig. \ref{fig_vir} (right). 
  We also include for comparison clouds observed in non-interacting galaxies and in the Galaxy. 
 
  Due to the big uncertainty in the virial parameter due to the $\alpha_{\rm{CO}}$ factor, the use of ionized gas instead of gas mass, 
 and the fact that to estimate this parameter we assume sphericity and homogeneous density,  
 for the virial parameter we do not find a tight correlation but a trend. The smaller the virial parameter, the higher is the $\Sigma_{\rm{SFR}}$ and the 
 $\Sigma_{\rm{mol}}$. Also, the smaller the virial parameter, the more gravitationally dominated is the cloud \citep{1992ApJ...395..140B}. Thus, 
 the massive clouds are more gravitationally bound and as a consequence 
 they have higher SFR and molecular gas surface densities.   
 The effect of being gravitationally bound is seen in the presence of the regions on the $L_{\rm{H\alpha}}-\sigma_v$ or $\Sigma_{\rm{SFR}}-\sigma_v$ envelopes.

\section{Discussion}

We present evidence here which supports the hypothesis that self-gravity is responsible for the enhancement of the molecular gas
and SFR densities in the most massive clouds, notably in LIRGs. In normal 
spirals, this enhancement is much less frequent \citep{zaragoza15}. \citet{gutierrez11} show that the HII regions are in
pressure equilibrium with the surrounding 
atomic gas in M51, where no enhancement of the H$\alpha$ luminosity was found (because the most luminous regions
do not reach the luminosities of the brightest regions in interacting galaxies), which implies no SFR enhancement. 
Since we see a trend between 
the virial parameter and the SFR-molecular gas surface density, we infer that this supports the scenario
in which self-gravity is enhancing the SFR. The presence 
of the $\rm{SFR}-\sigma_v$ envelope in HII regions shows that not all the regions have the same $\sigma_v$ driver. This 
can be seen by inspection of the H$\alpha$ surface brightness and velocity dispersion maps 
(Figs. \ref{fig_maps1}, \ref{fig_maps2} and \ref{fig_maps3} left). There, 
we see that the SFR is not simply correlated  with the velocity dispersion, although an envelope of values 
in H$\alpha$ correspond to
minima in the velocity dispersion. 
This effect is seen quantitatively in the $L_{\rm{H\alpha}}-\sigma_v$ plots from 
\citet{relano2005,2010MNRAS.407.2519B,zaragoza13,zaragoza14,zaragoza15} 
and in this work. 
The regions on the envelope show the minimum velocity dispersion for a given SFR,  
 while those far from the envelope have their $\sigma_v$ 
values enhanced. We can rule out  the widening 
of the H$\alpha$ emission profile by multiple kinematical components since we use the kinematical dimension to disentangle them. Thus, 
the $\sigma_v$ enhancement is likely due to  
 combined stellar feedback effects (winds and supernova). 
 Moreover, \citet{zaragoza14} showed that the HII regions on the envelope are those with enhanced SFR densities in the Antennae galaxies, 
which is an evidence of gravity domination in those HII regions.

The virial parameter presents values $\alpha_{\rm{vir}}>>1$ for clouds in equilibrium with external pressure while values $\alpha_{\rm{vir}}\sim1$ for clouds 
dominated by self gravity \citep{1992ApJ...395..140B}. We see a trend between the virial parameter, and the SFR and molecular gas surface densities, where the 
most intense star forming clouds have $\alpha_{\rm{vir}}$ closer to unity. Thus, they are more dominated by gravity rather than
external pressure. We infer that 
the feedback from the young ionizing stars does not dissolve the cloud if it is sufficiently massive and strongly bound, 
an effect which has been predicted theoretically \citep{2012MNRAS.424..377D,2016arXiv160600022H}.

Galaxy mergers produce a loss of axisymmetry, that induces gas flows towards the central parts of the galaxies 
\citep{1985AJ.....90..708K,1996ApJ...464..641M,2011EAS....51..107B}. Since LIRGs are frequently merging galaxies, these flows lead to an increase in the gas mass 
of the clouds in these systems, and when more mass is accreted into the clouds, they become more gravitationally bound and denser, enhancing the star formation.

 Due to the fact that the sizes of the clouds presented here are up to 460 pc for molecular clouds, and 900 pc for HII regions, one might think that the density enhancement is a resolution effect. 
 However, 
 the detection of massive star formation does not depend critically on the resolution, since as we can see in Fig. \ref{fig_dens} (left) 
there are more SFR enhanced HII regions in the Antennae galaxies compared to 
a sample of normal star forming galaxies from \citet{zaragoza15}, at similar resolution. 
 In addition it has been already observed that at similar resolutions HII regions in LIRGs are brighter than in normal galaxies 
\citep{2006ApJ...650..835A}. 
Also there is evidence that 
the molecular gas surface densities are higher at finer resolution \citep{2013ApJ...769L..12L,2016arXiv160607077L}, and that at smaller scales the clouds are more strongly gravitationally bound 
\citep{2016arXiv160607077L}. 
We can conclude that the higher molecular gas and SFR surface densities, and the gravitational bounding observed in these 
clouds in LIRGs are not due to the relatively low linear resolution.

 As we have shown in Fig. \ref{fig_dens}, the properties of the clouds in LIRGs form a bridge between those of the normal star forming clouds of nearby galaxies, 
and the most extreme cases observed at 
  $z\sim2$ and $z=3.4$. The masses of the clouds presented here are comparable with the masses of the clouds observed in massive star forming galaxies at $z>1$ \citep{2007ApJ...658..763E,ForsterSchreiber2011}. 
Thus, these nearby clouds are as massive and are forming stars at comparable rates to the clouds in massive star forming galaxies at $z>1$, and we can use them as  nearby analogs to understand what 
is driving, enhancing and quenching the star formation of galaxies in violent environments. 
Since the LIRGs are frequently interacting galaxies, we claim that the gas inflows induced by the interactions increase the density in the central zones, and in the overlap regions, and 
give rise to denser clouds with enhanced star formation efficiencies.  Galaxy mergers are more frequent at higher redshifts, while (U)LIRGs dominate the SFR density at $z>0.7$. Therefore, 
these self-gravity dominated star forming regions enhanced by mergers are a plausible driver of the higher SFRs in the past.

\section{Conclusions}

We have measured the properties of 88 HII regions and 27 molecular clouds in three systems of merging galaxies classified as LIRGs. We have compared the 
$\Sigma_{\rm{SFR}}$-$\rm{SFR}$ relation for the HII regions and the $\Sigma_{\rm{mol}}$-$M_{\rm{mol}}$ for the molecular clouds. For comparison we have used the clouds and HII regions of 
the Antennae galaxies (almost a LIRG) from \cite{zaragoza14}, and of normal star forming galaxies and the Galaxy \citep{2008ApJ...686..948B,2009ApJ...699.1092H,zaragoza15}. 
 We find that the $\Sigma_{\rm{SFR}}$-$\rm{SFR}$ relation is steeper for the HII regions in LIRGs compared to HII regions in normal star forming galaxies,  and 
that the $\Sigma_{\rm{mol}}$-$M_{\rm{mol}}$ trend seems steeper for the GMCs in LIRGS compared to those in normal star forming galaxies. However, due to the high dispersion of this relation, more 
quality data are needed to confirm the trend for GMCs. 
 Although in the case of HII regions H$\alpha$ intensity is affected by dust attenuation,  and for GMCs the $\alpha_{\rm{CO}}$ factor is not well constrained, the slopes of 
the relations should not change. The dust attenuation does not depend on the H$\alpha$ luminosity, so we expect to have a random attenuation over all the HII regions. 
The dependency of the $\alpha_{\rm{CO}}$ on the density, temperature, and metallicity makes also an under/over estimation of the molecular gas mass not affecting the trend in the 
$\Sigma_{\rm{mol}}$-$M_{\rm{mol}}$ relation.
This indicates that the 
SFR and the molecular gas surface densities are enhanced in these clouds, and they do not follow the classical third Larson's relation \citep{1981MNRAS.194..809L}, or the equivalent 
one for HII regions from \citet{1981MNRAS.195..839T}.  The properties of star forming regions in LIRGs form a bridge between the properties of those in normal star forming galaxies, and those 
at extreme star forming galaxies at higher redshifts.

The internal kinematics of the clouds is analysed through the velocity dispersion. For the HII regions, we confirm the previously reported $\rm{SFR}-\sigma_v$ envelope. The 
regions on the envelope are virialized because they are gravitationally dominated while the regions far from the envelope have enhanced velocity dispersion due to 
stellar feedback from winds and supernovae. The envelopes in the LIRGs and the normal star forming galaxies are similar, although \citet{zaragoza15} showed that interacting galaxies 
tend to have more HII regions on the envelope than non interacting galaxies. Since LIRGs are frequently interacting galaxies, and are forming stars at a higher rate, 
the HII regions in LIRGs should lie more frequently on the envelope, which is found. 

In the case of molecular clouds, we observe that the $\Sigma_{\rm{mol}}$-$\sigma_v$ relation seems to be steeper for the molecular clouds in the LIRGs than in the Galaxy and 
in normal non-interacting star forming galaxies. This indicates that molecular gas surface density is enhanced in the LIRGs clouds per unit of velocity dispersion. 
 The turbulence enhances the density of the molecular gas at scales 
roughly larger than $100\rm{pc}$. 
Since higher 
turbulence in LIRGs is found in denser clouds, and accompanies higher star formation rates, we suggest that this is because the clouds in the LIRGS are more dominated by self-gravity. In fact, we see that 
the lower values of the virial parameter in the HII regions and in molecular clouds correspond to denser clouds which drive enhanced star formation. 
 These turbulence enhanced star forming regions could be a plausible driver of higher SFRs observed at higher redshifts.

\section*{Acknowledgements}

JZC thanks the DGAPA Postdoctoral fellowships program of the 
National Autonomous University of Mexico (UNAM). 
MR and JZC acknowledge the grant IN103116 by DGAPA-PAPIIT UNAM.
AC and AB are supported by the programme of Residentes of the Instituto de Astrof\'isica de Canarias.
This work is partly supported by project P3/86 of the Instituto de Astrof\'isica de Canarias. JEB thanks the UNAM for partial support 
during a working visit. The authors thank Toshiki Saito for the reduced ALMA data of Arp 236. 
We really thank the anonymous referee, whose comments have led to important improvements on the original version of the paper.

Based on observations made with the William Herschel Telescope operated on the island of La Palma
by the Isaac Newton Group of Telescopes in the Spanish Observatorio del Roque de los
Muchachos of the
Instituto de Astrof\'isica de Canarias. 

This research made use of \textsc{Astropy}, a community-developed core Python package for Astronomy \citep{2013A&A...558A..33A},
\textsc{APLpy}, an open-source plotting package for Python hosted at http://aplpy.github.com, and \textsc{astrodendro},
a Python package to compute dendrograms of Astronomical data (http://www.dendrograms.org/).
The Atacama Large Millimeter/submillimeter Array (ALMA), an international astronomy facility,
is a partnership of Europe, North America and East Asia in cooperation with the Republic of Chile.
This paper makes use of the following ALMA data:
\newline
ADS/JAO.ALMA\#2011.0.00768.S  ADS/JAO.ALMA\#2011.0.00467.S  ADS/JAO.ALMA\#2013.1.00218.S





\begin{thebibliography}{99}

\bibitem[Allen(1973)]{1973asqu.book.....A} Allen, C.~W.\ 1973, London: 
University of London, Athlone Press, |c1973, 3rd ed.,  

\bibitem[Alonso-Herrero et al.(2006)]{2006ApJ...650..835A} Alonso-Herrero, A., Rieke, G.~H., Rieke, M.~J., et al.\ 2006, \apj, 650, 835 



\bibitem[Arribas et al.(2012)]{2012A&A...541A..20A} Arribas, S., Colina, L., Alonso-Herrero, A., et al.\ 2012, \aap, 541, A20 

\bibitem[Astropy Collaboration et
al.(2013)]{2013A&A...558A..33A} Astropy Collaboration, Robitaille, T.~P., Tollerud, E.~J., et al.\ 2013, \aap, 558, A33



\bibitem[Benn et al.(2008)]{2008SPIE.7014E..6XB} Benn, C., Dee, K., 
\& Ag{\'o}cs, T.\ 2008, Proceedings of the SPIE, 7014, 70146X 

%

\bibitem[Bertoldi
\& McKee(1992)]{1992ApJ...395..140B} Bertoldi, F., \& McKee, C.~F.\ 1992, \apj, 395, 140


\bibitem[Blasco-Herrera et al.(2010)]{2010MNRAS.407.2519B} Blasco-Herrera, 
J., Fathi, K., Beckman, J., et al.\ 2010, \mnras, 407, 2519 

\bibitem[\protect\citeauthoryear{Bolatto et
al.}{2008}]{2008ApJ...686..948B} Bolatto A.~D., Leroy A.~K., Rosolowsky E.,
Walter F., Blitz L., 2008, ApJ, 686, 948

\bibitem[Bolatto et al.(2013)]{2013ARA&A..51..207B} Bolatto, A.~D., Wolfire, M., \& Leroy, A.~K.\ 2013, \araa, 51, 207 


\bibitem[\protect\citeauthoryear{Bournaud}{2011}]{2011EAS....51..107B}
Bournaud F., 2011, EAS, 51, 107

\bibitem[Chapman et al.(2005)]{2005ApJ...622..772C} Chapman, S.~C., Blain, A.~W., Smail, I., \& Ivison, R.~J.\ 2005, \apj, 622, 772 

\bibitem[Ch{\'a}vez et al.(2014)]{chavez} Ch{\'a}vez, R., 
Terlevich, R., Terlevich, E., et al.\ 2014, \mnras, 442, 3565 


\bibitem[Daigle et al.(2006)]{2006MNRAS.368.1016D} Daigle, O., Carignan, 
C., Hernandez, O., Chemin, L., \& Amram, P.\ 2006, \mnras, 368, 1016 

\bibitem[Dale et al.(2012)]{2012MNRAS.424..377D} Dale, J.~E., Ercolano, B.,
\& Bonnell, I.~A.\ 2012, \mnras, 424, 377

\bibitem[Elbaz et al.(2007)]{2007A&A...468...33E} Elbaz, D., Daddi, E., Le Borgne, D., et al.\ 2007, \aap, 468, 33 

\bibitem[Elmegreen et al.(2007)]{2007ApJ...658..763E} Elmegreen, D.~M., Elmegreen, B.~G., Ravindranath, S., \& Coe, D.~A.\ 2007, \apj, 658, 763 



\bibitem[Erroz-Ferrer et al.(2012)]{2012MNRAS.427.2938E} Erroz-Ferrer, S.,
Knapen, J.~H., Font, J., et al.\ 2012, \mnras, 427, 2938

\bibitem[F{\"o}rster Schreiber et al.(2011)]{ForsterSchreiber2011} F{\"o}rster Schreiber, N.~M., Shapley, A.~E., Genzel, R., et al.\ 2011, \apj, 739, 45

\bibitem[Garc{\'{\i}}a-Mar{\'{\i}}n et al.(2006)]{2006ApJ...650..850G} Garc{\'{\i}}a-Mar{\'{\i}}n, M., Colina, L., Arribas, S., Alonso-Herrero, A., \& Mediavilla, E.\ 2006, \apj, 650, 850 



\bibitem[Green et al.(2014)]{2014MNRAS.437.1070G} Green, A.~W., Glazebrook, K., McGregor, P.~J., et al.\ 2014, \mnras, 437, 1070 

\bibitem[Guti{\'e}rrez et al.(2011)]{gutierrez11} Guti{\'e}rrez,
L., Beckman, J.~E., \& Buenrostro, V.\ 2011, \aj, 141, 113


\bibitem[Hernandez et al.(2008)]{2008PASP..120..665H} Hernandez, O., Fathi,
K., Carignan, C., et al.\ 2008, \pasp, 120, 665

\bibitem[Heyer et al.(2009)]{2009ApJ...699.1092H} Heyer, M., Krawczyk, C.,
Duval, J., \& Jackson, J.~M.\ 2009, \apj, 699, 1092

\bibitem[Howard et al.(2016)]{2016arXiv160600022H} Howard, C., Pudritz, R., \& Harris, W.\ 2016, arXiv:1606.00022 



\bibitem[Kartaltepe et al.(2010)]{2010ApJ...721...98K} Kartaltepe, J.~S., Sanders, D.~B., Le Floc'h, E., et al.\ 2010, \apj, 721, 98 



\bibitem[Keel et al.(1985)]{1985AJ.....90..708K} Keel, W.~C., Kennicutt, 
R.~C., Jr., Hummel, E., \& van der Hulst, J.~M.\ 1985, \aj, 90, 708 



\bibitem[Kennicutt(1998)]{1998ARA&A..36..189K} Kennicutt, R.~C., Jr.\ 1998, \araa, 36, 189 




\bibitem[Kim \& Sanders(1998)]{1998ApJS..119...41K} Kim, D.-C., \& Sanders, D.~B.\ 1998, \apjs, 119, 41 

\bibitem[Kormendy(2013)]{2013seg..book....1K} Kormendy, J.\ 2013, Secular Evolution of Galaxies, 1 

\bibitem[\protect\citeauthoryear{Larson}{1981}]{1981MNRAS.194..809L} Larson
R.~B., 1981, MNRAS, 194, 809

\bibitem[Le Floc'h et al.(2005)]{2005ApJ...632..169L} Le Floc'h, E., Papovich, C., Dole, H., et al.\ 2005, \apj, 632, 169 


\bibitem[Leroy et al.(2013)]{2013ApJ...769L..12L} Leroy, A.~K., Lee, C., Schruba, A., et al.\ 2013, \apjl, 769, L12 


\bibitem[Leroy et al.(2016)]{2016arXiv160607077L} Leroy, A.~K., Hughes, A., Schruba, A., et al.\ 2016, arXiv:1606.07077 

\bibitem[Ma et al.(2016)]{2016arXiv160908553M} Ma, J., Gonzalez, A.~H., Vieira, J.~D., et al.\ 2016, arXiv:1609.08553 


\bibitem[Magnelli et al.(2011)]{magnelli11} Magnelli, B., Elbaz, D., Chary, R.~R., et al.\ 2011, \aap, 528, A35 


\bibitem[Magnelli et al.(2013)]{magnelli13} Magnelli, B., Popesso, P., Berta, S., et al.\ 2013, \aap, 553, A132 

\bibitem[Micha{\l}owski et al.(2010)]{2010A&A...514A..67M} Micha{\l}owski, M., Hjorth, J., \& Watson, D.\ 2010, \aap, 514, A67 


\bibitem[Mihos 
\& Hernquist(1996)]{1996ApJ...464..641M} Mihos, J.~C., \& Hernquist, L.\ 1996, \apj, 464, 641 






\bibitem[Oteo et al.(2016)]{2016arXiv160706464O} Oteo, I., Zwaan, M.~A., Ivison, R.~J., Smail, I., \& Biggs, A.~D.\ 2016, arXiv:1607.06464 

\bibitem[Papadopoulos et al.(2012)]{2012ApJ...751...10P} Papadopoulos, P.~P., van der Werf, P., Xilouris, E., Isaak, K.~G., \& Gao, Y.\ 2012, \apj, 751, 10 

\bibitem[P{\'e}rez-Gonz{\'a}lez et al.(2005)]{2005ApJ...630...82P} P{\'e}rez-Gonz{\'a}lez, P.~G., Rieke, G.~H., Egami, E., et al.\ 2005, \apj, 630, 82 

\bibitem[Piqueras L{\'o}pez et al.(2016)]{2016A&A...590A..67P} Piqueras L{\'o}pez, J., Colina, L., Arribas, S., Pereira-Santaella, M., \& Alonso-Herrero, A.\ 2016, \aap, 590, A67 

\bibitem[Rela{\~n}o et al.(2005)]{relano2005} Rela{\~n}o, M., Beckman, J.~E., Zurita, A., Rozas, M., \& Giammanco, C.\ 2005, \aap, 431, 235 

\bibitem[Roman-Duval et al.(2010)]{2010ApJ...723..492R} Roman-Duval, J.,
Jackson, J.~M., Heyer, M., Rathborne, J.,
\& Simon, R.\ 2010, \apj, 723, 492


\bibitem[Rodighiero et al.(2011)]{2011ApJ...739L..40R} Rodighiero, G., Daddi, E., Baronchelli, I., et al.\ 2011, \apjl, 739, L40 

\bibitem[Rodr{\'{\i}}guez-Zaur{\'{\i}}n et al.(2011)]{2011A&A...527A..60R} Rodr{\'{\i}}guez-Zaur{\'{\i}}n, J., Arribas, S., Monreal-Ibero, A., et al.\ 2011, \aap, 527, A60 

\bibitem[Rosolowsky
\& Leroy(2006)]{2006PASP..118..590R} Rosolowsky, E., \& Leroy, A.\ 2006, \pasp, 118, 590

\bibitem[Rosolowsky et al.(2008)]{2008ApJ...679.1338R} Rosolowsky, E.~W.,
Pineda, J.~E., Kauffmann, J., \& Goodman, A.~A.\ 2008, \apj, 679, 1338

%
\bibitem[Saito et al.(2015)]{2015ApJ...803...60S} Saito, T., Iono, D., Yun, M.~S., et al.\ 2015, \apj, 803, 60 



\bibitem[Sanders et al.(2003)]{2003AJ....126.1607S} Sanders, D.~B.,
Mazzarella, J.~M., Kim, D.-C., Surace, J.~A.,
\& Soifer, B.~T.\ 2003, \aj, 126, 1607

\bibitem[\protect\citeauthoryear{Schweizer et
al.}{2008}]{2008AJ....136.1482S} Schweizer F., et al., 2008, AJ, 136, 1482

\bibitem[Sliwa et al.(2014)]{2014ApJ...796L..15S} Sliwa, K., Wilson, C.~D., Iono, D., Peck, A., \& Matsushita, S.\ 2014, \apjl, 796, L15 

\bibitem[Solomon et al.(1992)]{1992ApJ...398L..29S} Solomon, P.~M., Downes,
D., \& Radford, S.~J.~E.\ 1992, \apjl, 398, L29

\bibitem[\protect\citeauthoryear{Spitzer}{1978}]{1978spitzer}
Spitzer L., 1978, Physical processes in the interstellar medium, by Lyman Spitzer.~ New York Wiley-Interscience, 1978.~333 p.

\bibitem[Springel et al.(2005)]{2005Natur.435..629S} Springel, V., White, S.~D.~M., Jenkins, A., et al.\ 2005, \nat, 435, 629 

\bibitem[Tacconi et al.(2008)]{2008ApJ...680..246T} Tacconi, L.~J., Genzel, R., Smail, I., et al.\ 2008, \apj, 680, 246-262 

\bibitem[Terlevich\& Melnick(1981)]{1981MNRAS.195..839T} Terlevich, R., \& Melnick, J.\ 1981, \mnras, 195, 839

\bibitem[Terlevich et al.(2015)]{2015MNRAS.451.3001T} Terlevich, R., Terlevich, E., Melnick, J., et al.\ 2015, \mnras, 451, 3001 

\bibitem[Ueda et al.(2012)]{2012ApJ...745...65U} Ueda, J., Iono, D., 
Petitpas, G., et al.\ 2012, \apj, 745, 65 



\bibitem[Wei et al.(2012)]{2012ApJ...750..136W} Wei, L.~H., Keto, E.,
\& Ho, L.~C.\ 2012, \apj, 750, 136

\bibitem[Wuyts et al.(2011)]{2011ApJ...742...96W} Wuyts, S., F{\"o}rster Schreiber, N.~M., van der Wel, A., et al.\ 2011, \apj, 742, 96 



\bibitem[Zaragoza-Cardiel et al.(2013)]{zaragoza13} Zaragoza-Cardiel, J., Font-Serra, J., Beckman, J.~E., et al.\ 2013, \mnras, 432, 998 

\bibitem[Zaragoza-Cardiel et al.(2014)]{zaragoza14} 
Zaragoza-Cardiel, J., Font, J., Beckman, J.~E., et al.\ 2014, \mnras, 445, 1412 

\bibitem[Zaragoza-Cardiel et al.(2015)]{zaragoza15} Zaragoza-Cardiel, J., Beckman, J.~E., Font, J., et al.\ 2015, \mnras, 451, 1307 


\end{thebibliography}




\appendix

\section{Properties of star forming regions tables}
\begin{table*}

 \scriptsize
\centering
 \begin{minipage}{140mm}

  \caption{Physical properties of the HII regions derived as described in section $\S3$. 
The table 
is available as a machine readable table in the electronic version of the paper and through CDS. }
  \begin{tabular}{@{}ccccccccccc@{}}

  \hline
 Object &$\rm{N_{cl}}$  &RA&Dec & $\log(L_{\rm{H\alpha}}) $ & $\log(\mathrm{SFR})$& $R_{\rm{HII}}$ &  $\log(\Sigma_{\rm{SFR}})$ & $\log(M_{\rm{HII}})$ &$\sigma_{v}$& $\alpha_{vir}$ \\
 & &     (hh:mm:ss)         &     ($^{\circ}$ $\mathrm{\prime}$ $\mathrm{\prime\prime}$) &  ($\mathrm{erg/s}$) & ($\mathrm{M_{\odot}/\mathrm{yr}}$)  &   (pc)   & ($\mathrm{\frac{M_{\odot}}{yr\thinspace kpc^2}}$)&($\mathrm{M_{\odot}}$) &         ($\, \mathrm{km/s}$)    & \\                              \\
  \hline
\\
Arp 236 & 1 & 1: 7:46.8 & -17:30:30.2 & $ 37.88\pm0.05 $ & $ -3.22\pm0.05 $ & $ 170\pm70 $ & $ -2.2\pm0.8 $ & $ 5.3\pm0.6 $ &  $<4$  &  \\ 
Arp 236 & 2 & 1: 7:46.5 & -17:30:25.8 & $ 37.92\pm0.06 $ & $ -3.18\pm0.06 $ & $ 130\pm70 $ &  $ -2\pm1 $ & $ 5.2\pm0.8 $ &  $<4$  &  \\ 
Arp 236 & 3 & 1: 7:46.2 & -17:30:31.5 & $ 38.0\pm0.1 $ & $ -3.1\pm0.1 $ & $ 70\pm30 $ & $ -1.4\pm0.8 $ & $ 4.8\pm0.6 $ &  $<4$  &  \\ 
Arp 236 & 4 & 1: 7:46.9 & -17:30:25.4 & $ 38.02\pm0.05 $ & $ -3.09\pm0.05 $ & $ 110\pm40 $ & $ -1.7\pm0.8 $ & $ 5.1\pm0.6 $ &  $<4$  &  \\ 
Arp 236 & 5 & 1: 7:47.0 & -17:30:25.3 & $ 38.09\pm0.06 $ & $ -3.01\pm0.06 $ & $ 160\pm40 $ & $ -1.9\pm0.6 $ & $ 5.4\pm0.4 $ &  $<4$  &  \\ 
Arp 236 & 6 & 1: 7:47.1 & -17:30:24.2 & $ 38.15\pm0.03 $ & $ -2.96\pm0.03 $ & $ 140\pm40 $ & $ -1.7\pm0.6 $ & $ 5.3\pm0.4 $ &  $<4$  &  \\ 
Arp 236 & 7 & 1: 7:46.9 & -17:30:30.3 & $ 38.18\pm0.03 $ & $ -2.93\pm0.03 $ & $ 210\pm30 $ & $ -2.1\pm0.4 $ & $ 5.6\pm0.3 $ &  $<4$  &  \\ 
Arp 236 & 8 & 1: 7:46.6 & -17:30:15.5 & $ 38.2\pm0.02 $ & $ -2.9\pm0.02 $ & $ 80\pm30 $ & $ -1.2\pm0.7 $ & $ 5.0\pm0.5 $ &  $<4$  &  \\ 
Arp 236 & 9 & 1: 7:46.7 & -17:30:27.7 & $ 38.27\pm0.02 $ & $ -2.83\pm0.02 $ & $ 180\pm40 $ & $ -1.8\pm0.4 $ & $ 5.5\pm0.3 $ &  $<4$  &  \\ 
Arp 236 & 10 & 1: 7:46.7 & -17:30:36.4 & $ 38.29\pm0.02 $ & $ -2.81\pm0.02 $ & $ 150\pm50 $ & $ -1.6\pm0.6 $ & $ 5.4\pm0.5 $ &  $<4$  &  \\ 
Arp 236 & 11 & 1: 7:46.9 & -17:30:38.8 & $ 38.31\pm0.02 $ & $ -2.8\pm0.02 $ & $ 140\pm50 $ & $ -1.6\pm0.8 $ & $ 5.4\pm0.6 $ &  $<4$  &  \\ 
Arp 236 & 12 & 1: 7:46.7 & -17:30:31.6 & $ 38.34\pm0.05 $ & $ -2.76\pm0.05 $ & $ 310\pm80 $ & $ -2.2\pm0.6 $ & $ 5.9\pm0.4 $ &  $<4$  &  \\ 
Arp 236 & 13 & 1: 7:46.9 & -17:30:29.0 & $ 38.43\pm0.02 $ & $ -2.67\pm0.02 $ & $ 350\pm30 $ & $ -2.3\pm0.2 $ & $ 6.1\pm0.1 $ &  $<4$  &  \\ 
Arp 236 & 14 & 1: 7:46.5 & -17:30:21.6 & $ 38.48\pm0.05 $ & $ -2.62\pm0.05 $ & $ 110\pm20 $ & $ -1.2\pm0.5 $ & $ 5.3\pm0.4 $ &  $<4$  &  \\ 
Arp 236 & 15 & 1: 7:46.6 & -17:30:22.2 & $ 38.521\pm0.008 $ & $ -2.582\pm0.008 $ & $ 150\pm50 $ & $ -1.4\pm0.7 $ & $ 5.6\pm0.5 $ &  $<4$  &  \\ 
Arp 236 & 16 & 1: 7:46.3 & -17:30:32.5 & $ 38.58\pm0.06 $ & $ -2.52\pm0.06 $ & $ 220\pm30 $ & $ -1.7\pm0.3 $ & $ 5.8\pm0.2 $ &  $<4$  &  \\ 
Arp 236 & 17 & 1: 7:46.3 & -17:30:29.6 & $ 38.879\pm0.006 $ & $ -2.223\pm0.006 $ & $ 210\pm40 $ & $ -1.4\pm0.4 $ & $ 6.0\pm0.3 $ &  $<4$  &  \\ 
Arp 236 & 18 & 1: 7:47.2 & -17:30:28.0 & $ 38.89\pm0.02 $ & $ -2.21\pm0.02 $ & $ 270\pm70 $ & $ -1.6\pm0.5 $ & $ 6.1\pm0.4 $ &  $<4$  &  \\ 
Arp 236 & 19 & 1: 7:46.3 & -17:30:28.7 & $ 38.91\pm0.01 $ & $ -2.19\pm0.01 $ & $ 60\pm20 $ & $ -0.3\pm0.6 $ & $ 5.2\pm0.5 $ &   $ 7\pm5 $ & $ 30\pm60 $ \\ 
Arp 236 & 20 & 1: 7:46.4 & -17:30:28.7 & $ 38.946\pm0.006 $ & $ -2.157\pm0.006 $ & $ 160\pm30 $ & $ -1.1\pm0.4 $ & $ 5.8\pm0.3 $ &  $<4$  &  \\ 
Arp 236 & 21 & 1: 7:47.3 & -17:30:26.8 & $ 38.95\pm0.01 $ & $ -2.15\pm0.01 $ & $ 80\pm20 $ & $ -0.4\pm0.4 $ & $ 5.3\pm0.3 $ &  $<4$  &  \\ 
Arp 236 & 22 & 1: 7:46.3 & -17:30:25.1 & $ 38.961\pm0.007 $ & $ -2.142\pm0.007 $ & $ 240\pm10 $ & $ -1.4\pm0.1 $ & $ 6.07\pm0.09 $ &  $ 12\pm3 $ & $ 30\pm20 $ \\ 
Arp 236 & 23 & 1: 7:46.3 & -17:30:28.0 & $ 38.971\pm0.005 $ & $ -2.132\pm0.005 $ & $ 410\pm30 $ & $ -1.9\pm0.2 $ & $ 6.4\pm0.1 $ &  $ 10\pm3 $ & $ 20\pm10 $ \\ 
Arp 236 & 24 & 1: 7:46.9 & -17:30:24.1 & $ 38.984\pm0.009 $ & $ -2.119\pm0.009 $ & $ 130\pm30 $ & $ -0.8\pm0.5 $ & $ 5.7\pm0.4 $ &  $<4$  &  \\ 
Arp 236 & 25 & 1: 7:46.3 & -17:30:29.5 & $ 38.985\pm0.007 $ & $ -2.118\pm0.007 $ & $ 340\pm70 $ & $ -1.7\pm0.4 $ & $ 6.3\pm0.3 $ &  $<4$  &  \\ 
Arp 236 & 26 & 1: 7:46.3 & -17:30:29.7 & $ 39.02\pm0.01 $ & $ -2.08\pm0.01 $ & $ 170\pm60 $ & $ -1.0\pm0.7 $ & $ 5.9\pm0.5 $ &  $<4$  &  \\ 
Arp 236 & 27 & 1: 7:46.7 & -17:30:23.7 & $ 39.18\pm0.04 $ & $ -1.92\pm0.04 $ & $ 440\pm40 $ & $ -1.7\pm0.2 $ & $ 6.6\pm0.2 $ &  $<4$  &  \\ 
Arp 236 & 28 & 1: 7:46.7 & -17:30:25.9 & $ 39.21\pm0.05 $ & $ -1.9\pm0.05 $ & $ 130\pm40 $ & $ -0.6\pm0.6 $ & $ 5.8\pm0.4 $ &  $ 23\pm6 $ & $ 100\pm200 $ \\ 
Arp 236 & 29 & 1: 7:46.9 & -17:30:23.9 & $ 39.25\pm0.02 $ & $ -1.86\pm0.02 $ & $ 360\pm50 $ & $ -1.5\pm0.3 $ & $ 6.5\pm0.2 $ &   $ 5\pm3 $ &   $ 4\pm5 $ \\ 
Arp 236 & 30 & 1: 7:47.6 & -17:30:23.9 & $ 39.26\pm0.02 $ & $ -1.84\pm0.02 $ & $ 210\pm70 $ & $ -1.0\pm0.7 $ & $ 6.1\pm0.5 $ &  $<4$  &  \\ 
Arp 236 & 31 & 1: 7:47.5 & -17:30:24.6 & $ 39.274\pm0.008 $ & $ -1.828\pm0.008 $ & $ 210\pm50 $ & $ -1.0\pm0.5 $ & $ 6.1\pm0.4 $ &  $<4$  &  \\ 
Arp 236 & 32 & 1: 7:47.1 & -17:30:23.0 & $ 39.3\pm0.01 $ & $ -1.8\pm0.01 $ & $ 220\pm30 $ & $ -1.0\pm0.3 $ & $ 6.2\pm0.2 $ &  $<4$  &  \\ 
Arp 236 & 33 & 1: 7:47.5 & -17:30:19.9 & $ 39.347\pm0.007 $ & $ -1.755\pm0.007 $ & $ 250\pm30 $ & $ -1.1\pm0.2 $ & $ 6.3\pm0.2 $ &   $ 9\pm2 $ &  $ 11\pm7 $ \\ 
Arp 236 & 34 & 1: 7:47.2 & -17:30:26.6 & $ 39.354\pm0.007 $ & $ -1.749\pm0.007 $ & $ 180\pm40 $ & $ -0.8\pm0.4 $ & $ 6.1\pm0.3 $ &  $<4$  &  \\ 
Arp 236 & 35 & 1: 7:46.5 & -17:30:21.0 & $ 39.452\pm0.006 $ & $ -1.65\pm0.006 $ & $ 420\pm30 $ & $ -1.4\pm0.2 $ & $ 6.7\pm0.1 $ &  $<4$  &  \\ 
Arp 236 & 36 & 1: 7:46.6 & -17:30:19.3 & $ 39.476\pm0.007 $ & $ -1.626\pm0.007 $ & $ 280\pm30 $ & $ -1.0\pm0.2 $ & $ 6.4\pm0.2 $ &  $<4$  &  \\ 
Arp 236 & 37 & 1: 7:46.9 & -17:30:21.4 & $ 39.517\pm0.004 $ & $ -1.585\pm0.004 $ & $ 200\pm30 $ & $ -0.7\pm0.3 $ & $ 6.2\pm0.2 $ &  $ 10\pm3 $ & $ 10\pm10 $ \\ 
Arp 236 & 38 & 1: 7:46.5 & -17:30:23.2 & $ 39.679\pm0.009 $ & $ -1.423\pm0.009 $ & $ 450\pm30 $ & $ -1.2\pm0.1 $ & $ 6.8\pm0.1 $ &  $<4$  &  \\ 
Arp 236 & 39 & 1: 7:47.1 & -17:30:16.5 & $ 39.789\pm0.008 $ & $ -1.314\pm0.008 $ & $ 420\pm20 $ & $ -1.06\pm0.08 $ & $ 6.86\pm0.06 $ &  $<4$  &  \\ 
Arp 236 & 40 & 1: 7:46.7 & -17:30:26.4 & $ 40.14\pm0.006 $ & $ -0.963\pm0.006 $ & $ 620\pm10 $ & $ -1.04\pm0.05 $ & $ 7.28\pm0.04 $ &  $ 15\pm1 $ &   $ 8\pm2 $ \\ 
Arp 236 & 41 & 1: 7:46.6 & -17:30:25.3 & $ 40.16\pm0.04 $ & $ -0.94\pm0.04 $ & $ 170\pm10 $ & $ 0.1\pm0.2 $ & $ 6.5\pm0.1 $ &  $ 32\pm2 $ & $ 70\pm20 $ \\ 
Arp 236 & 42 & 1: 7:46.5 & -17:30:25.1 & $ 40.296\pm0.006 $ & $ -0.807\pm0.006 $ & $ 550\pm10 $ & $ -0.79\pm0.05 $ & $ 7.29\pm0.03 $ &  $ 22\pm2 $ &  $ 16\pm3 $ \\ 
Arp 236 & 43 & 1: 7:46.5 & -17:30:24.8 & $ 40.531\pm0.004 $ & $ -0.571\pm0.004 $ & $ 450\pm10 $ & $ -0.37\pm0.07 $ & $ 7.27\pm0.05 $ &   $ 7\pm1 $ & $ 1.5\pm0.7 $ \\ 
Arp 236 & 44 & 1: 7:47.1 & -17:30:23.6 & $ 41.087\pm0.005 $ & $ -0.015\pm0.005 $ & $ 550\pm20 $ & $ 0.0\pm0.06 $ & $ 7.68\pm0.05 $ &  $ 26\pm1 $ &   $ 9\pm2 $ \\ 
Arp 236 & 45 & 1: 7:46.7 & -17:30:26.4 & $ 41.21\pm0.008 $ & $ 0.108\pm0.008 $ & $ 920\pm10 $ & $ -0.31\pm0.03 $ & $ 8.07\pm0.02 $ & $ 25.8\pm0.8 $ & $ 6.0\pm0.6 $ \\ 
Arp 236 & 46 & 1: 7:46.5 & -17:30:21.1 & $ 41.692\pm0.007 $ & $ 0.59\pm0.007 $ & $ 890\pm20 $ & $ 0.19\pm0.04 $ & $ 8.3\pm0.03 $ & $ 24.9\pm0.9 $ & $ 3.3\pm0.4 $ \\ 
Arp 186 & 1 & 4:33:59.8 &  -8:34:45.6 & $ 38.22\pm0.08 $ & $ -2.88\pm0.08 $ & $ 130\pm30 $ & $ -1.6\pm0.5 $ & $ 5.3\pm0.3 $ &  $<8$  &  \\ 
Arp 186 & 2 & 4:34: 0.1 &  -8:34:45.3 & $ 38.24\pm0.04 $ & $ -2.87\pm0.04 $ & $ 90\pm40 $ & $ -1.3\pm0.8 $ & $ 5.1\pm0.6 $ &  $<8$  &  \\ 
Arp 186 & 3 & 4:34: 0.2 &  -8:34:45.9 & $ 38.24\pm0.03 $ & $ -2.86\pm0.03 $ & $ 180\pm50 $ & $ -1.9\pm0.6 $ & $ 5.5\pm0.5 $ &  $<8$  &  \\ 
Arp 186 & 4 & 4:33:60.0 &  -8:34:45.5 & $ 38.47\pm0.02 $ & $ -2.64\pm0.02 $ & $ 130\pm20 $ & $ -1.3\pm0.3 $ & $ 5.4\pm0.2 $ &  $<8$  &  \\ 
Arp 186 & 5 & 4:34: 0.2 &  -8:34:50.8 & $ 38.56\pm0.03 $ & $ -2.54\pm0.03 $ & $ 120\pm40 $ & $ -1.2\pm0.6 $ & $ 5.4\pm0.5 $ &  $<8$  &  \\ 
Arp 186 & 6 & 4:33:59.5 &  -8:34:49.4 & $ 38.6\pm0.1 $ & $ -2.5\pm0.1 $ & $ 160\pm40 $ & $ -1.4\pm0.5 $ & $ 5.6\pm0.4 $ &  $<8$  &  \\ 
Arp 186 & 7 & 4:33:59.8 &  -8:34:51.8 & $ 38.61\pm0.06 $ & $ -2.49\pm0.06 $ &  $ 52\pm4 $ & $ -0.4\pm0.2 $ & $ 4.9\pm0.1 $ & $ 40\pm10 $ & $ 1000\pm1000 $ \\ 
Arp 186 & 8 & 4:34: 0.4 &  -8:34:46.1 & $ 38.63\pm0.02 $ & $ -2.47\pm0.02 $ & $ 130\pm30 $ & $ -1.2\pm0.4 $ & $ 5.5\pm0.3 $ &  $ 21\pm9 $ & $ 200\pm300 $ \\ 
Arp 186 & 9 & 4:33:59.6 &  -8:34:53.6 & $ 38.7\pm0.1 $ & $ -2.4\pm0.1 $ & $ 140\pm30 $ & $ -1.2\pm0.6 $ & $ 5.6\pm0.4 $ &  $<8$  &  \\ 
Arp 186 & 10 & 4:33:59.7 &  -8:34:50.4 & $ 38.68\pm0.02 $ & $ -2.42\pm0.02 $ & $ 210\pm30 $ & $ -1.6\pm0.3 $ & $ 5.8\pm0.2 $ &  $ 27\pm5 $ & $ 200\pm200 $ \\ 
Arp 186 & 11 & 4:33:59.5 &  -8:34:49.1 & $ 38.7\pm0.02 $ & $ -2.4\pm0.02 $ & $ 160\pm30 $ & $ -1.3\pm0.4 $ & $ 5.7\pm0.3 $ &  $<8$  &  \\ 
Arp 186 & 12 & 4:34: 1.4 &  -8:35: 2.9 & $ 38.82\pm0.01 $ & $ -2.29\pm0.01 $ & $ 210\pm40 $ & $ -1.4\pm0.4 $ & $ 5.9\pm0.3 $ &  $ 13\pm8 $ & $ 50\pm90 $ \\ 
Arp 186 & 13 & 4:33:59.6 &  -8:34:47.1 & $ 38.909\pm0.008 $ & $ -2.194\pm0.008 $ & $ 170\pm40 $ & $ -1.2\pm0.5 $ & $ 5.8\pm0.3 $ &  $<8$  &  \\ 
Arp 186 & 14 & 4:33:59.4 &  -8:34:56.3 & $ 39.0\pm0.01 $ & $ -2.1\pm0.01 $ & $ 120\pm20 $ & $ -0.7\pm0.4 $ & $ 5.6\pm0.3 $ &  $<8$  &  \\ 
Arp 186 & 15 & 4:34: 0.5 &  -8:34:43.4 & $ 39.11\pm0.005 $ & $ -1.992\pm0.005 $ & $ 100\pm30 $ & $ -0.5\pm0.5 $ & $ 5.6\pm0.4 $ &  $<8$  &  \\ 
Arp 186 & 16 & 4:34: 0.6 &  -8:34:41.8 & $ 39.136\pm0.006 $ & $ -1.967\pm0.006 $ & $ 320\pm40 $ & $ -1.5\pm0.2 $ & $ 6.4\pm0.2 $ &  $<8$  &  \\ 
Arp 186 & 17 & 4:33:59.5 &  -8:34:50.8 & $ 39.35\pm0.01 $ & $ -1.75\pm0.01 $ & $ 170\pm40 $ & $ -0.7\pm0.4 $ & $ 6.1\pm0.3 $ &  $<8$  &  \\ 
Arp 186 & 18 & 4:33:59.3 &  -8:35: 2.5 & $ 39.677\pm0.005 $ & $ -1.425\pm0.005 $ & $ 220\pm40 $ & $ -0.6\pm0.4 $ & $ 6.4\pm0.3 $ &  $<8$  &  \\ 
Arp 186 & 19 & 4:33:59.7 &  -8:34:44.7 & $ 39.91\pm0.01 $ & $ -1.19\pm0.01 $ & $ 550\pm40 $ & $ -1.2\pm0.1 $ & $ 7.1\pm0.1 $ &   $ 9\pm1 $ &   $ 4\pm2 $ \\ 
Arp 186 & 20 & 4:34: 0.2 &  -8:34:43.7 & $ 39.96\pm0.02 $ & $ -1.14\pm0.02 $ & $ 300\pm20 $ & $ -0.6\pm0.2 $ & $ 6.7\pm0.1 $ &  $ 47\pm4 $ & $ 150\pm50 $ \\ 
Arp 186 & 21 & 4:34: 0.1 &  -8:34:50.1 & $ 40.1\pm0.01 $ & $ -1.0\pm0.01 $ & $ 580\pm20 $ & $ -1.0\pm0.1 $ & $ 7.22\pm0.07 $ & $ 14.6\pm0.8 $ &   $ 9\pm2 $ \\ 
Arp 186 & 22 & 4:33:60.0 &  -8:34:50.2 & $ 40.77\pm0.01 $ & $ -0.34\pm0.01 $ & $ 420\pm20 $ & $ -0.1\pm0.1 $ & $ 7.34\pm0.07 $ &  $ 16\pm2 $ &   $ 5\pm2 $ \\ 
Arp 186 & 23 & 4:33:59.8 &  -8:34:43.9 & $ 41.923\pm0.009 $ & $ 0.821\pm0.009 $ & $ 740\pm10 $ & $ 0.58\pm0.04 $ & $ 8.29\pm0.03 $ &  $ 73\pm1 $ &  $ 24\pm2 $ \\ 

  \end{tabular}
    \end{minipage}
 \label{table_hii}
\end{table*}

\begin{table*}

 \scriptsize
\centering
 \begin{minipage}{140mm}

  \contcaption{}
  \begin{tabular}{@{}ccccccccccc@{}}

  \hline
 Object &$\rm{N_{cl}}$  &RA&Dec & $\log(L_{\rm{H\alpha}}) $ & $\log(\mathrm{SFR})$& $R_{\rm{HII}}$ &  $\log(\Sigma_{\rm{SFR}})$ & $\log(M_{\rm{HII}})$ &$\sigma_{v}$& $\alpha_{vir}$ \\
 & &     (hh:mm:ss)         &     ($^{\circ}$ $\mathrm{\prime}$ $\mathrm{\prime\prime}$) &  ($\mathrm{erg/s}$) & ($\mathrm{M_{\odot}/\mathrm{yr}}$)  &   (pc)   & ($\mathrm{\frac{M_{\odot}}{yr\thinspace kpc^2}}$)&($\mathrm{M_{\odot}}$) &         ($\, \mathrm{km/s}$)    & \\                              \\
  \hline
\\
Arp 298 & 1 &23: 3:17.3 &   8:53:34.2 & $ 39.3\pm0.03 $ & $ -1.81\pm0.03 $ & $ 220\pm20 $ & $ -1.0\pm0.2 $ & $ 6.2\pm0.2 $ &  $<4$  &  \\ 
Arp 298 & 2 &23: 3:17.2 &   8:53:45.6 & $ 39.38\pm0.02 $ & $ -1.73\pm0.02 $ & $ 260\pm20 $ & $ -1.1\pm0.2 $ & $ 6.3\pm0.2 $ &  $<4$  &  \\ 
Arp 298 & 3 &23: 3:17.4 &   8:53:41.2 & $ 39.41\pm0.01 $ & $ -1.69\pm0.01 $ & $ 170\pm10 $ & $ -0.6\pm0.2 $ & $ 6.1\pm0.1 $ &  $ 52\pm5 $ & $ 400\pm200 $ \\ 
Arp 298 & 4 &23: 3:17.6 &   8:53:37.9 & $ 39.419\pm0.009 $ & $ -1.683\pm0.009 $ & $ 130\pm30 $ & $ -0.4\pm0.5 $ & $ 5.9\pm0.4 $ &  $<4$  &  \\ 
Arp 298 & 5 &23: 3:17.9 &   8:53:37.3 & $ 39.48\pm0.02 $ & $ -1.62\pm0.02 $ & $ 260\pm20 $ & $ -1.0\pm0.1 $ & $ 6.4\pm0.1 $ &   $ 6\pm1 $ &   $ 4\pm3 $ \\ 
Arp 298 & 6 &23: 3:17.7 &   8:53:40.1 & $ 39.5\pm0.01 $ & $ -1.6\pm0.01 $ & $ 290\pm30 $ & $ -1.0\pm0.2 $ & $ 6.5\pm0.1 $ &  $ 12\pm2 $ & $ 20\pm10 $ \\ 
Arp 298 & 7 &23: 3:16.2 &   8:52:21.1 & $ 39.52\pm0.01 $ & $ -1.59\pm0.01 $ & $ 310\pm20 $ & $ -1.1\pm0.1 $ & $ 6.52\pm0.08 $ &   $ 9\pm1 $ &   $ 9\pm3 $ \\ 
Arp 298 & 8 &23: 3:15.6 &   8:52:24.6 & $ 39.6\pm0.01 $ & $ -1.5\pm0.01 $ & $ 320\pm20 $ & $ -1.0\pm0.1 $ & $ 6.59\pm0.09 $ &   $ 4\pm1 $ &   $ 2\pm1 $ \\ 
Arp 298 & 9 &23: 3:15.6 &   8:52:18.3 & $ 39.67\pm0.01 $ & $ -1.44\pm0.01 $ & $ 310\pm20 $ & $ -0.9\pm0.1 $ & $ 6.6\pm0.08 $ &   $ 5\pm1 $ &   $ 2\pm1 $ \\ 
Arp 298 & 10 &23: 3:16.0 &   8:52:26.7 & $ 39.69\pm0.01 $ & $ -1.41\pm0.01 $ & $ 120\pm20 $ & $ -0.1\pm0.3 $ & $ 6.0\pm0.2 $ &  $<4$  &  \\ 
Arp 298 & 11 &23: 3:18.2 &   8:53:37.0 & $ 39.744\pm0.008 $ & $ -1.359\pm0.008 $ & $ 380\pm20 $ & $ -1.0\pm0.1 $ & $ 6.77\pm0.08 $ &  $ 13\pm2 $ &  $ 14\pm6 $ \\ 
Arp 298 & 12 &23: 3:16.6 &   8:52:49.5 & $ 39.758\pm0.005 $ & $ -1.345\pm0.005 $ & $ 140\pm20 $ & $ -0.1\pm0.4 $ & $ 6.1\pm0.3 $ &  $<4$  &  \\ 
Arp 298 & 13 &23: 3:15.6 &   8:52:27.1 & $ 39.782\pm0.007 $ & $ -1.32\pm0.007 $ & $ 280\pm10 $ & $ -0.7\pm0.1 $ & $ 6.58\pm0.07 $ &  $ 23\pm1 $ & $ 50\pm10 $ \\ 
Arp 298 & 14 &23: 3:15.3 &   8:52:24.6 & $ 39.94\pm0.02 $ & $ -1.16\pm0.02 $ & $ 320\pm10 $ & $ -0.66\pm0.08 $ & $ 6.75\pm0.06 $ &   $ 8\pm1 $ &   $ 4\pm2 $ \\ 
Arp 298 & 15 &23: 3:14.7 &   8:52:56.4 & $ 40.0\pm0.01 $ & $ -1.11\pm0.01 $ & $ 370\pm20 $ & $ -0.7\pm0.1 $ & $ 6.87\pm0.09 $ & $ 10.4\pm0.9 $ &   $ 6\pm2 $ \\ 
Arp 298 & 16 &23: 3:15.5 &   8:52:25.7 & $ 40.04\pm0.01 $ & $ -1.07\pm0.01 $ & $ 410\pm20 $ & $ -0.8\pm0.1 $ & $ 6.97\pm0.07 $ &  $ 22\pm1 $ &  $ 25\pm6 $ \\ 
Arp 298 & 17 &23: 3:14.9 &   8:52:54.1 & $ 40.08\pm0.02 $ & $ -1.02\pm0.02 $ & $ 560\pm40 $ & $ -1.0\pm0.1 $ & $ 7.2\pm0.1 $ &  $<4$  &  \\ 
Arp 298 & 18 &23: 3:15.3 &   8:52:42.4 & $ 40.39\pm0.01 $ & $ -0.71\pm0.01 $ & $ 210\pm20 $ & $ 0.1\pm0.2 $ & $ 6.7\pm0.2 $ &  $ 16\pm3 $ &  $ 12\pm9 $ \\ 
Arp 298 & 19 &23: 3:15.6 &   8:52:27.1 & $ 40.47\pm0.02 $ & $ -0.63\pm0.02 $ & $ 460\pm20 $ & $ -0.46\pm0.09 $ & $ 7.26\pm0.06 $ &  $ 18\pm1 $ &   $ 9\pm2 $ \\ 

  \end{tabular}
    \end{minipage}
 \label{table_hii}
\end{table*}

\begin{table*}
 \scriptsize
\centering
 \begin{minipage}{140mm}

  \caption{Physical properties of the molecular clouds derived as described in section $\S3$. 
The table 
is available as a machine readable table in the electronic version of the paper and through CDS. }
  \begin{tabular}{@{}cccccccccc@{}}

  \hline
 Object &$\rm{N_{cl}}$  &RA&Dec & $\log(L_{\rm{CO}}) $ & $\log(M_{\mathrm{mol}})$& $R_{\rm{mol}}$ &  $\log(\Sigma_{\rm{mol}})$ & $\sigma_{v}$& $\alpha_{vir}$ \\
 & &     (hh:mm:ss)         &     ($^{\circ}$ $\mathrm{\prime}$ $\mathrm{\prime\prime}$) &  ($\mathrm{K\thinspace km\thinspace s^{-1}\thinspace pc^2}$)
 & ($\mathrm{M_{\odot}}$)  &   (pc)   & ($\mathrm{\frac{M_{\odot}}{pc^2}}$) &         ($\, \mathrm{km/s}$)    & \\                              \\
  \hline
\\
Arp 236 & 1 & 1: 7:47.1 & -17:30:26.7 & $ 8.045\pm0.005 $ & $ 9.115\pm0.005 $ & $ 348\pm4 $ & $ 3.53\pm0.03 $ &  $ 30\pm1 $ & $ 0.28\pm0.02 $ \\ 
Arp 236 & 2 & 1: 7:47.2 & -17:30:23.7 & $ 7.181\pm0.007 $ & $ 8.251\pm0.007 $ & $ 230\pm7 $ & $ 3.03\pm0.07 $ &  $ 21\pm2 $ & $ 0.6\pm0.1 $ \\ 
Arp 236 & 3 & 1: 7:47.5 & -17:30:25.6 & $ 8.38\pm0.01 $ & $ 9.45\pm0.01 $ & $ 388\pm6 $ & $ 3.78\pm0.04 $ & $ 27.8\pm0.7 $ & $ 0.12\pm0.01 $ \\ 
Arp 236 & 4 & 1: 7:46.8 & -17:30:24.6 & $ 6.52\pm0.01 $ & $ 7.59\pm0.01 $ & $ 180\pm20 $ & $ 2.6\pm0.2 $ & $ 7.3\pm0.7 $ & $ 0.29\pm0.08 $ \\ 
Arp 236 & 5 & 1: 7:46.8 & -17:30:25.4 & $ 7.801\pm0.006 $ & $ 8.871\pm0.006 $ & $ 460\pm10 $ & $ 3.05\pm0.05 $ &  $ 24\pm1 $ & $ 0.41\pm0.05 $ \\ 
Arp 236 & 6 & 1: 7:47.4 & -17:30:26.1 & $ 7.281\pm0.003 $ & $ 8.351\pm0.003 $ & $ 225\pm7 $ & $ 3.15\pm0.07 $ & $ 9.9\pm0.5 $ & $ 0.11\pm0.01 $ \\ 
Arp 186 & 1 & 4:34: 0.0 &  -8:34:46.0 & $ 7.45\pm0.002 $ & $ 8.52\pm0.002 $ & $ 124\pm2 $ & $ 3.83\pm0.04 $ & $ 18.1\pm0.3 $ & $ 0.144\pm0.008 $ \\ 
Arp 186 & 2 & 4:33:59.9 &  -8:34:47.4 & $ 6.962\pm0.006 $ & $ 8.032\pm0.006 $ & $ 187\pm3 $ & $ 2.99\pm0.04 $ & $ 11.8\pm0.2 $ & $ 0.28\pm0.02 $ \\ 
Arp 186 & 3 & 4:34: 0.1 &  -8:34:48.0 & $ 6.707\pm0.005 $ & $ 7.777\pm0.005 $ & $ 141\pm4 $ & $ 2.98\pm0.06 $ & $ 8.7\pm0.2 $ & $ 0.21\pm0.02 $ \\ 
Arp 186 & 4 & 4:34: 0.1 &  -8:34:47.2 & $ 6.12\pm0.01 $ & $ 7.19\pm0.01 $ &  $ 85\pm3 $ & $ 2.83\pm0.09 $ &  $< 5\rm{km/s}$ &  \\ 
Arp 186 & 5 & 4:33:59.9 &  -8:34:45.9 & $ 6.728\pm0.003 $ & $ 7.798\pm0.003 $ & $ 106\pm2 $ & $ 3.25\pm0.04 $ & $ 14.1\pm0.5 $ & $ 0.39\pm0.03 $ \\ 
Arp 186 & 6 & 4:34: 0.1 &  -8:34:45.3 & $ 6.327\pm0.005 $ & $ 7.397\pm0.005 $ &  $ 45\pm2 $ & $ 3.59\pm0.08 $ & $ 13.4\pm0.9 $ & $ 0.38\pm0.06 $ \\ 
Arp 186 & 7 & 4:33:60.0 &  -8:34:45.3 & $ 7.062\pm0.004 $ & $ 8.132\pm0.004 $ & $ 76.7\pm0.9 $ & $ 3.87\pm0.03 $ & $ 16.0\pm0.8 $ & $ 0.17\pm0.02 $ \\ 
Arp 186 & 8 & 4:34: 0.1 &  -8:34:43.7 & $ 6.959\pm0.004 $ & $ 8.029\pm0.004 $ & $ 153\pm2 $ & $ 3.16\pm0.03 $ & $ 12.1\pm0.3 $ & $ 0.24\pm0.02 $ \\ 
Arp 186 & 9 & 4:34: 0.2 &  -8:34:42.2 & $ 6.658\pm0.008 $ & $ 7.728\pm0.008 $ & $ 112\pm2 $ & $ 3.13\pm0.04 $ & $ 13.9\pm0.3 $ & $ 0.47\pm0.03 $ \\ 
Arp 186 & 10 & 4:34: 0.0 &  -8:34:44.2 & $ 7.362\pm0.003 $ & $ 8.432\pm0.003 $ & $ 127\pm3 $ & $ 3.73\pm0.05 $ & $ 18.4\pm0.4 $ & $ 0.18\pm0.01 $ \\ 
Arp 186 & 11 & 4:34: 0.1 &  -8:34:42.3 & $ 6.198\pm0.006 $ & $ 7.268\pm0.006 $ &  $ 56\pm2 $ & $ 3.27\pm0.07 $ & $ 14.5\pm0.7 $ & $ 0.7\pm0.1 $ \\ 
Arp 186 & 12 & 4:34: 0.1 &  -8:34:43.6 & $ 6.702\pm0.003 $ & $ 7.772\pm0.003 $ &  $ 91\pm1 $ & $ 3.35\pm0.03 $ & $ 19.8\pm0.5 $ & $ 0.7\pm0.05 $ \\ 
Arp 298 & 1 &23: 3:15.7 &   8:52:26.2 & $ 7.382\pm0.006 $ & $ 8.197\pm0.006 $ & $ 210\pm10 $ & $ 3.1\pm0.1 $ &  $ 26\pm2 $ & $ 1.1\pm0.2 $ \\ 
Arp 298 & 2 &23: 3:15.6 &   8:52:21.2 & $ 6.74\pm0.02 $ & $ 7.56\pm0.02 $ & $ 170\pm10 $ & $ 2.6\pm0.1 $ & $ 9.7\pm0.9 $ & $ 0.5\pm0.1 $ \\ 
Arp 298 & 3 &23: 3:16.0 &   8:52:29.0 & $ 6.98\pm0.01 $ & $ 7.8\pm0.01 $ & $ 216\pm7 $ & $ 2.63\pm0.07 $ & $ 17.9\pm0.8 $ & $ 1.3\pm0.2 $ \\ 
Arp 298 & 4 &23: 3:15.6 &   8:52:24.2 & $ 6.78\pm0.01 $ & $ 7.6\pm0.01 $ &  $ 98\pm7 $ & $ 3.1\pm0.2 $ &   $ 9\pm1 $ & $ 0.24\pm0.08 $ \\ 
Arp 298 & 5 &23: 3:15.5 &   8:52:25.0 & $ 7.47\pm0.01 $ & $ 8.29\pm0.01 $ & $ 161\pm8 $ & $ 3.4\pm0.1 $ &  $ 17\pm1 $ & $ 0.26\pm0.05 $ \\ 
Arp 298 & 6 &23: 3:15.7 &   8:52:29.1 & $ 6.8\pm0.01 $ & $ 7.61\pm0.01 $ & $ 134\pm9 $ & $ 2.9\pm0.2 $ & $ 10.5\pm0.7 $ & $ 0.42\pm0.09 $ \\ 
Arp 298 & 7 &23: 3:15.3 &   8:52:25.8 & $ 6.53\pm0.02 $ & $ 7.35\pm0.02 $ & $ 130\pm10 $ & $ 2.6\pm0.2 $ & $ 5.7\pm0.5 $ & $ 0.23\pm0.06 $ \\ 
Arp 298 & 8 &23: 3:15.5 &   8:52:27.1 & $ 7.825\pm0.004 $ & $ 8.639\pm0.004 $ & $ 357\pm6 $ & $ 3.04\pm0.04 $ & $ 11.4\pm0.5 $ & $ 0.12\pm0.01 $ \\ 
Arp 298 & 9 &23: 3:15.4 &   8:52:26.0 & $ 6.9\pm0.01 $ & $ 7.71\pm0.01 $ & $ 190\pm10 $ & $ 2.6\pm0.1 $ & $ 9.2\pm0.2 $ & $ 0.37\pm0.04 $ \\ 

  \end{tabular}
    \end{minipage}
 \label{table_mol}
\end{table*}


\bsp	
\label{lastpage}
\end{document}